\documentclass[final,11pt]{article}
\usepackage{multirow}
\usepackage[nottoc]{tocbibind}
\usepackage{geometry}
\usepackage{subcaption}
\usepackage{placeins}
\usepackage{ulem}
\usepackage[affil-it, auth-sc]{authblk}
\usepackage[english]{babel}
\usepackage[utf8]{inputenc}
\usepackage{graphicx}
\usepackage{amssymb}
\usepackage{amsthm}
\usepackage{amsmath}
\usepackage{amsfonts}
\usepackage{verbatim}
\usepackage{lineno}
\usepackage{todonotes}
\presetkeys{todonotes}{inline, color=blue!30, size=\small}{}
\usepackage{color, soul}
\definecolor{darkred}{rgb}{0.9, 0.0, 0.0}
\definecolor{darkgreen}{rgb}{0.0, 0.5, 0.0}
\definecolor{fnalblue}{RGB}{ 0, 76, 151}
\definecolor{fnalbluedark}{RGB}{  0, 40, 85}
\usepackage[colorlinks,bookmarks,
linkcolor=fnalbluedark,
citecolor=darkgreen
,urlcolor={blue!50!black}
]{hyperref}
\usepackage{eso-pic}
\usepackage[sort&compress,numbers]{natbib}
\usepackage{doi}
\usepackage{paralist,epsfig}
\usepackage{relsize}
\usepackage{nicefrac,esint}
\usepackage{float}
\usepackage{cleveref}

\newcommand{\nl}{\nonumber \\ }
\def\fAt{f_{A3}}
\def\fAtbar{\bar{f}_{A3}}

\def\FVone{F^V_1}
\def\FVtwo{F^V_2}
\def\FSone{F^S_1}
\def\FStwo{F^S_2}

\begin{document}

\AddToShipoutPictureFG*{\AtPageUpperLeft{\put(-60,-75){\makebox[\paperwidth][r]{FERMILAB-PUB-24-0531-T,~LA-UR-22-22625,~LLNL-JRNL-2014635}}}}

\title{\Large\bf Nucleon axial-vector form factor and radius from radiatively-corrected antineutrino scattering data}

\author[1,2]{Oleksandr~Tomalak}
\affil[1]{Institute of Theoretical Physics, Chinese Academy of Sciences, Beijing 100190, P. R. China \vspace{1.2mm}}
\affil[2]{Theoretical Division, Los Alamos National Laboratory, Los Alamos, NM 87545, USA \vspace{1.2mm}}

\author[3]{Aaron~S.~Meyer}
\affil[3]{Nuclear \& Chemical Sciences Division, Lawrence Livermore National Laboratory, Livermore, CA 94550, USA\vspace{1.2mm}}

\author[4]{Clarence~Wret}
\affil[4]{Department of Physics, Imperial College London, London, SW7 2BW, United Kingdom\vspace{1.2mm}}

\author[5,6]{Tejin~Cai}
\affil[5]{Department of Physics and Astronomy, University of Rochester, Rochester, NY 14627, USA\vspace{1.2mm}}
\affil[6]{Department of Physics and Astronomy, York University, Toronto, ONT M3J 1P3, Canada\vspace{1.2mm}}

\author[7,8]{Richard~J.~Hill}
\affil[7]{Department of Physics and Astronomy, University of Kentucky, Lexington, KY 40506, USA\vspace{1.2mm}}
\affil[8]{Fermilab, Batavia, IL 60510, USA\vspace{1.2mm}}

\author[5]{Kevin~S.~McFarland}

\date{\today}

\maketitle
\begin{abstract}
The nucleon axial-vector form factor, $G_A$, is critical to determine the electroweak interactions of leptons with nucleons. Important examples of processes influenced by $G_A$ are elastic (anti)neutrino-nucleon scattering and muon capture by the proton. Sparse experimental data results in a large uncertainty on the momentum dependence of $G_A$ and has motivated the consideration of new experimental probes and first-principles lattice quantum chromodynamics (QCD) evaluations. The comparison of new and precise theoretical predictions for $G_A$ with future experimental data necessitates the application of radiative corrections to experimentally-observable processes. We apply these corrections in the extraction of $G_A$ and the associated axial-vector radius from the recent MINERvA antineutrino-hydrogen data, compare the effects from radiative corrections to other uncertainties in neutrino scattering experiments, and discuss the comparison of lattice QCD evaluations to experimental measurements.
\end{abstract}

\vspace*{2cm}
\begin{center}
    \includegraphics[width=0.15\textwidth]{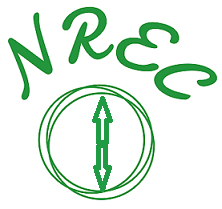} 
\end{center}
\begin{center}
    \vspace{-3mm}
    \tiny{\textbf{Nuclear Radius \\ Extraction Collaboration}}
\end{center}

\newpage
\tableofcontents
\newpage

\section{Introduction}

The charged-current (anti)neutrino-nucleon elastic scattering is, at leading order in quantum electrodynamics, described by four structure-dependent nucleon form factors. These form factors have been the subject of active experimental and theoretical research during the last $70$ years. Assuming isospin symmetry, two of these form factors---electric and magnetic---describe electromagnetic interactions with nucleons and have been probed at the percent level and below from electron scattering experiments and atomic spectroscopy. Electroweak interactions with neutrinos require two more form factors: the axial-vector and the induced pseudoscalar. At GeV scale neutrino energies, the nucleon axial-vector form factor is the dominant source of the uncertainty in the unpolarized (anti)neutrino-nucleon cross sections. Although the normalization of the axial-vector form factor is known precisely from neutron beta decay measurements, the momentum dependence of this form factor is much less precisely known.

The traditional database for the nucleon axial-vector form factor includes neutrino scattering data and pion electroproduction measurements. Experiments measuring the scattering of neutrinos on nucleons were performed in the deuterium bubble chambers at ANL~\cite{Mann:1973pr,Barish:1977qk,Miller:1982qi}, Gargamelle~\cite{Bonetti:1977cs}, BNL~\cite{Baker:1981su,Kitagaki:1990vs}, FNAL~\cite{Kitagaki:1983px}, and BEBC~\cite{Allasia:1990uy}, and scattering of antineutrinos on the hydrogen atoms inside hydrocarbons was recently measured by MINERvA~\cite{MINERvA:2023avz}. Pion electroproduction experiments~\cite{Amaldi:1970tg,Amaldi:1972vf,Bloom:1973fn,Brauel:1973cw,Joos:1976ng,DelGuerra:1975uiy,DelGuerra:1976uj,Esaulov:1978ed,Choi:1993vt,A1:1999kwj,A1:2016emg} allow us to access the axial-vector radius (cf. Eq.~(\ref{eq:rAdef}) below) with measurements at relatively low momentum transfers, where the higher terms in the squared momentum transfer expansion---which are specific to pion production---do not contribute significantly.

Even with the large uncertainties on the data, the constraints on the form factors from the deuterium bubble chamber experiments~\cite{Meyer:2016oeg} are in tension with the antineutrino-hydrogen scattering data~\cite{MINERvA:2023avz,Tomalak:2023pdi,Tomalak:2024yvq,Meyer:2025rzh}. Recent lattice quantum chromodynamics (LQCD) calculations~\cite{Djukanovic:2022wru,RQCD:2019jai,Park:2021ypf,Alexandrou:2020okk,Jang:2023zts} are consistent with the antineutrino-hydrogen scattering data~\cite{Tomalak:2023pdi,Meyer:2025rzh,Meyer:2026kdl} but not with the deuterium data~\cite{Meyer:2025rzh,Meyer:2026kdl}. With promising sub-percent projections for future neutrino experiments~\cite{Petti:2023abz} and comparable precision goals for cross sections in neutrino oscillation experiments~\cite{Hyper-KamiokandeProto-:2015xww,DUNE:2020jqi}, these discrepancies motivate further improvements in the precision of lattice QCD calculations.

To provide a consistent theoretical description of the elastic (anti)neutrino-nucleon scattering cross sections at the percent level and below, we develop and apply the framework of radiative corrections for the extraction of the nucleon axial-vector form factor and radius from the (anti)neutrino scattering data for the first time. After applying the quantum electrodynamics (QED) radiative corrections from Refs.~\cite{Tomalak:2021hec,Tomalak:2022xup}, with currently negligible uncertainties~\cite{Tomalak:2025vxa}, we extract the axial-vector form factor from the antineutrino-hydrogen scattering data of MINERvA~\cite{MINERvA:2023avz} and evaluate the corresponding axial-vector radius. We study the effects of accounting for the radiative corrections and updating the nucleon vector form factors with recent data from A1$@$MAMI~\cite{Bernauer:2010wm,Bernauer:2013tpr,Borah:2020gte} in the extraction of the nucleon axial-vector form factor and radius from the MINERvA data as well as from pseudodata that corresponds to the neutrino fluxes of the DUNE, Hyper-K, MINERvA, and BEBC experiments.

The remainder of the paper is organized as follows. In Section~\ref{sec2}, we present the formalism of scattering amplitudes, define the nucleon axial-vector form factor and radius in the presence of radiative corrections, and relate the form-factor normalizations to the experimental values of the nucleon isovector-vector and axial-vector coupling constants. We describe $z$-expansion functional form of the fit for the nucleon axial-vector form factor in Section~\ref{sec3}. In Section~\ref{sec4}, we outline details of QED radiative corrections, which we apply in the analysis of the experimental data. We present the fit results, with the nucleon axial-vector form factor and radius for the MINERvA antineutrino data and study the effects from radiative corrections for the neutrino pseudodata in Section~\ref{sec5}. In Section~\ref{sec6}, we discuss future precision of experiments that will be able extract the nucleon axial-vector form factor and radius. We highlight future directions to achieve a percent-level precision with LQCD simulations in Section~\ref{sec7}. Our conclusions and outlook are presented in Section~\ref{sec8}. We discuss the choice of the fit parameters in Appendices~\ref{app:t0_choice} and~\ref{app:t0_variation} and provide fit results with a larger number of significant digits in the Supplementary Material.

\section{Nucleon axial-vector form factor and radius with radiative corrections}
\label{sec2}

In this Section, we define the nucleon axial-vector form factor and radius in the presence of radiative corrections. We provide process-independent definitions by specifying the renormalization scheme as in Refs.~\cite{Tomalak:2021hec,Tomalak:2022xup}.

First, we present the general Lorentz-invariant decomposition of the elastic (anti)neutrino-nucleon scattering matrix elements, which is valid including the electromagnetic radiative corrections. For massless leptons, the charged-current elastic (anti)neutrino-nucleon matrix elements $T^{m_\ell = 0}_{\nu_\ell n \to \ell^- p}$ and $T^{m_\ell = 0}_{\bar{\nu}_\ell p \to \ell^+ n}$ are described by $4$ invariant amplitudes $f_1,~f_2,~f_A$, and~$f_{A3}$ as~\cite{Tomalak:2021hec,Tomalak:2022xup}\footnote{We use the shorthand notation $\bar{\ell}^- (\dots ) \nu_{\ell} = \bar{u}^{(\ell)}(p^\prime) (\dots ) u^{(\nu)}(p)$ and $\overline{\bar{\nu}}_\ell (\dots) \ell^+= \bar{v}^{(\nu)}(p^\prime) (\dots ) v^{(\ell)}(p)$ for the usual Dirac spinors.}
\begin{align}
T^{m_\ell = 0}_{\nu_\ell n \to \ell^- p} &= \sqrt{2}\mathrm{G}_\mathrm{F} V_{ud} \, \bar{\ell}^- \gamma^\mu \mathrm{P}_\mathrm{L} \nu_{\ell}\, \bar{p} \left(\gamma_\mu \left( f_1 + f_2 + {f}_A \gamma_5 \right) - \left( f_2 - 2 \fAt \gamma_5 \right) \frac{{K}_\mu}{M} \right) n \,, \nl
T^{m_\ell = 0}_{\bar{\nu}_\ell p \to \ell^+ n} &= \sqrt{2}\mathrm{G}_\mathrm{F} V^*_{ud} \, \overline{\bar{\nu}}_\ell \gamma^\mu \mathrm{P}_\mathrm{L} \ell^+ \, \bar{n} \left( \gamma_\mu \left( \bar{f}_1 + \bar{f}_2 + \bar{f}_A \gamma_5 \right) - \left( \bar{f}_2 + 2 \fAtbar \gamma_5 \right) \frac{{K}_\mu}{M} \right) p , \label{eq:CCQE_amplitude}
\end{align}
with the Fermi coupling constant $\mathrm{G}_F$, the Cabibbo-Kobayashi-Maskawa matrix element $V_{ud}$, the nucleon mass $M$, and the averaged nucleon momentum $K_\mu = (k_\mu + k^\prime_\mu)/2$. After accounting for the charged lepton mass $m_\ell$, four additional amplitudes $f_3,~f_{P},~f_{R}$, and $f_T$ contribute to the elastic scattering~\cite{Tomalak:2024yvq,Borah:2024hvo}:
\begin{align}
T^{m_\ell \neq 0}_{\nu_\ell n \to \ell^- p} &= \sqrt{2}\mathrm{G}_\mathrm{F} V_{ud} \frac{m_\ell}{M} \left[\frac{{f}_{T}}{4} \, \bar{\ell}^- \sigma^{\mu \nu} \mathrm{P}_\mathrm{L} \nu_{\ell}\, \bar{p} \sigma_{\mu \nu} n -\bar{\ell}^- \mathrm{P}_\mathrm{L} \nu_{\ell}\, \bar{p} \left( {f}_{3} + f_P \gamma_5 - \frac{f_R}{4} \frac{\gamma.P}{M} \gamma_5 \right) n \right]\,, \nl
T^{m_\ell \neq 0}_{\bar{\nu}_\ell p \to \ell^+ n} &=\sqrt{2}\mathrm{G}_\mathrm{F} V^*_{ud} \frac{m_\ell}{M} \left[ \frac{\bar{f}_{T}}{4} \, \overline{\bar{\nu}}_\ell \sigma^{\mu \nu} \mathrm{P}_\mathrm{R} \ell^+ \, \bar{n} \sigma_{\mu \nu} p - \overline{\bar{\nu}}_\ell \mathrm{P}_\mathrm{R} \ell^+ \, \bar{n} \left( \bar{f}_{3} + \bar{f}_P \gamma_5 - \frac{\bar{f}_R}{4} \frac{\gamma.P}{M} \gamma_5 \right) p \right], \,\label{eq:CCQE_amplitudem}
\end{align} 
with the averaged lepton momentum $P_\mu = (p_\mu + p^\prime_\mu)/2$, projectors on the left-handed and right-handed chiral states $\mathrm{P}_\mathrm{L} = \left(1-\gamma_5 \right)/2$ and $\mathrm{P}_\mathrm{R} = \left(1+\gamma_5 \right)/2$, respectively. The complete amplitudes are then given by $T_{\nu_\ell n \to \ell^- p} = T^{m_\ell = 0}_{\nu_\ell n \to \ell^- p} + T^{m_\ell \neq 0}_{\nu_\ell n \to \ell^- p}$, and $T_{\bar{\nu}_\ell p \to \ell^+ n} = T^{m_\ell = 0}_{\bar{\nu}_\ell p \to \ell^+ n} + T^{m_\ell \neq 0}_{\bar{\nu}_\ell p \to \ell^+ n}$. We note that the amplitudes in neutrino and antineutrino scattering are connected by the crossing relation $\bar{f}^j_i \left( \nu+i0, Q^2 \right) = f^j_i \left( -\nu-i0, Q^2 \right)^*$, where $\nu = E_\nu / M - \tau - r_\ell^2$ with the (anti)neutrino energy $E_\nu$, $\tau = Q^2/(4M^2)$, $Q^2 = - \left( p - p^\prime \right)^2$, and $r_\ell = m_\ell/\left( 2M \right)$. At leading order, the amplitudes $f_1,~f_2,~f_A$, and $f_P$ are nonzero and correspond to the nucleon form factors. Radiative corrections contribute to all eight invariant amplitudes. Denoting the virtual and soft radiative corrections with superscript $v$, we express the amplitudes $f_1,~f_2,~f_A$, and $f_P$ in terms of the ``Born" form factors $F_{V1},~F_{V2},~G_{A}$, and $F_{P}$, respectively,~\cite{Tomalak:2022xup}
\begin{align}
 f_1 (\nu, Q^2) &= \sqrt{Z_\ell Z_h^{(p)}} \left( F_{V1}(Q^2) + f^v_1 (\nu, Q^2) \right)\,, \nl
 f_2 (\nu, Q^2) &= \sqrt{Z_\ell Z_h^{(p)}} \left(F_{V2}(Q^2) + f^v_2 (\nu, Q^2) \right)\,, \nl
 f_A (\nu, Q^2) &= \sqrt{Z_\ell Z_h^{(p)}} \left( -G_A(Q^2) + f^v_A (\nu, Q^2)\right)\,, \nl
 f_P (\nu, Q^2) &= \sqrt{Z_\ell Z_h^{(p)}} \left(F_P(Q^2) +f^v_P (\nu, Q^2) \right) \,, \label{eq:hadronic_model}
\end{align}
while other amplitudes have only virtual contributions, i.e., $f^j_i (\nu, Q^2) = \sqrt{Z_\ell Z_h^{(p)}} \left(f^j_i \right)^v (\nu, Q^2)$ for $f_3,~f_{A3},~f_{R}$, and $f_T$. In Eqs.~(\ref{eq:hadronic_model}), we denote the field renormalization factors for the relativistic charged lepton as $Z_\ell$ and for the heavy proton as $Z_h^{(p)}$. We specify the definition of Born form factors in Eqs.~(\ref{eq:hadronic_model}) in the $\overline{\rm MS}$ renormalization scheme with renormalization scale $\mu=M$~\cite{Hill:2016gdf} and gauge-dependent parameter $\xi_\gamma = 1$, i.e., the radiative corrections are evaluated in the Feynman-'t Hooft gauge~\cite{Tomalak:2021hec,Tomalak:2022xup} at the scale of the nucleon mass.

At low energies, the nucleon axial-vector coupling constant $g^\mathrm{exp}_A$ is extracted from measurements of the asymmetry in polarized neutron beta decay~\cite{UCNA:2017obv,Markisch:2018ndu,Dubbers:2018kgh}, where it enters in the ratio to the vector coupling constant $g_V$ as $g^\mathrm{exp}_A/g_V$.\footnote{The precise relation beyond the leading order in $\alpha$ between $g^\mathrm{exp}_A$, and lattice QCD evaluations~\cite{RQCD:2019jai,Bali:2023sdi,Alexandrou:2020okk,Djukanovic:2022wru,Tsuji:2022ric,Park:2021ypf,Jang:2023zts} of the matrix element of the isovector axial-vector quark current between the nucleon states, $\langle N^\prime | \overline{u} \gamma_\mu \gamma_5 u - \overline{d} \gamma_\mu \gamma_5 d | N\rangle$, with subsequent extrapolation to the forward kinematics, is under active investigation~\cite{Cirigliano:2022hob,Cirigliano:2023fnz,Cirigliano:2024nfi,Tomalak:2026wks}.} Having precisely determined $g_V$ from the Standard Model~\cite{Cirigliano:2023fnz}, the experimental value for $g^\mathrm{exp}_A$ and the theoretical value for $g_V$ at the chiral scale $\mu_\chi = m_e$ can be matched to the four-fermion amplitudes $f_A$ and $f_1$, respectively, after accounting for the radiative corrections in the low-energy pionless effective field theory.\footnote{For a discussion of $\overline{\rm MS}_\chi$ used in Ref.~\cite{Cirigliano:2023fnz} versus conventional $\overline{\rm MS}$, see e.g. Ref.~\cite{Cao:2025pjt}.} Using the hadronic model of Ref.~\cite{Tomalak:2021hec,Tomalak:2022xup} for $f_A$ and $f_1$, we thus  relate the normalization of the axial-vector form factor $G_A \left( 0 \right)$ and the isovector-vector form factor $F_{V 1} \left( 0 \right)$ to the PDG-quoted value $g^\mathrm{exp}_A$~\cite{ParticleDataGroup:2022pth} and $g_V$:
\begin{align}
    G_A \left( 0 \right)  &\; = \;  g_A \left( \mu_\chi = m_e \right) \left( 1  - \frac{\alpha}{4 \pi} \frac{2}{4-d}  + \frac{\alpha}{8 \pi} \ln \frac{m_e^2}{M^2} - \frac{3\alpha}{8 \pi} + \frac{\alpha}{\pi} \int \hspace{-0.1cm} \frac{i \mathrm{d}^d L}{ \left( 2 \pi \right)^{d-2}} \frac{m_e^{4-d} \left[ \FVone \left(q^2\right) - 1 \right]}{L^2 \left( L^2 - \lambda^2_\gamma \right)}  \right) \nl
    &+ \frac{e^2}{3} \int \hspace{-0.1cm} \frac{i \mathrm{d}^4 L}{ \left( 2 \pi \right)^4} \frac{2 F_{V1} \left( 0 \right) \left( 2 \FSone \left(q^2\right) + \FStwo\left(q^2\right) \right) + G_{A} \left( 0 \right) \left( 4 \FVtwo\left(q^2\right) + 5 \FVone \left(q^2\right)\right)}{L^2 \left( \left( L + k\right)^2 - M^2\right)} \nl
    &- \frac{e^2}{3} \int \hspace{-0.1cm} \frac{i \mathrm{d}^4 L}{ \left( 2 \pi \right)^4} \frac{F_{V1} \left( 0 \right) \left( \FSone \left(q^2\right) + 2 \FStwo\left(q^2\right) \right) - G_{A} \left( 0 \right) \left( \FVone\left(q^2\right) + \FVtwo \left(q^2\right) / 2 \right)}{L^2 \left(\left( L + k\right)^2 - M^2 \right)} \frac{k \cdot L}{M^2}, \\
    F_{V1} \left( 0 \right)  &\; = \;  g_V \left( \mu_\chi = m_e \right) \left( 1  - \frac{\alpha}{4 \pi} \frac{2}{4-d}  + \frac{\alpha}{8 \pi} \ln \frac{m_e^2}{M^2} - \frac{3\alpha}{8 \pi} + \frac{\alpha}{\pi} \int \hspace{-0.1cm} \frac{i \mathrm{d}^d L}{ \left( 2 \pi \right)^{d-2}} \frac{m_e^{4-d} \left[ \FVone \left(q^2\right) - 1 \right]}{L^2 \left( L^2 - \lambda^2_\gamma \right)} \right) \nl
    &+ e^2 \int \hspace{-0.1cm} \frac{i \mathrm{d}^4 L}{ \left( 2 \pi \right)^4}  \frac{2 G_A \left( 0 \right) \left( \FSone \left(q^2\right) + \FStwo\left(q^2\right) \right) +  F_{V1} \left( 0 \right) \FVone \left(q^2\right) }{L^2 \left( \left( L + k\right)^2 - M^2\right)} \nl
    &+ e^2  \int \hspace{-0.1cm} \frac{i \mathrm{d}^4 L}{ \left( 2 \pi \right)^4} \frac{G_A \left( 0 \right) \left( \FSone \left(q^2\right) + \FStwo\left(q^2\right) \right) - F_{V1} \left( 0 \right) \left( \FVone \left(q^2\right) + 3 \FVtwo \left(q^2\right) / 2 \right)}{L^2 \left(\left( L + k\right)^2 - M^2 \right)} \frac{k \cdot L}{M^2},
\end{align}
with $q^2 = - L^2$, and $d\to 4$ dimensions of the space-time.  We use the notations $F^V_{i} = F^p_i - F^n_i$ and $F^S_{i} = F^p_i + F^n_i$ for the isovector-vector and isoscalar-vector electromagnetic form factors, respectively, with proton $F^p_i$ and neutron $F^n_i$ form factors. Numerically, the matching condition for the radiative corrections from Refs.~\cite{Tomalak:2021hec,Tomalak:2022xup} is\footnote{For tree-level fits, we take the PDG value of the axial-vector $g_A$ over the vector $g_V$ coupling constants ratio $\lambda^\mathrm{exp} = g_A/g_V = 1.2754$ as a normalization, i.e., $G_A \left( Q^2 = 0 \right) = \lambda^\mathrm{exp}$ and $F_{V1} \left( Q^2 = 0 \right) = 1$. For fits with the radiative corrections, we account for the Standard Model short-distance contributions to the vector coupling constant from Ref.~\cite{Cirigliano:2023fnz}: $g_V = 1.02499(13)$. Accounting also for the decrease of the PDG value for the nucleon axial-vector coupling constant from $g^\mathrm{exp}_A = 1.3041(23)$~\cite{ParticleDataGroup:2014cgo,Meyer:2016oeg} to $g^\mathrm{exp}_A = \lambda^\mathrm{exp} g_V = 1.3073(13)$~\cite{ParticleDataGroup:2022pth} after the PERKEO-III measurement~\cite{Markisch:2018ndu}, we normalize the axial-vector form factor in fits, which account for the radiative corrections, as $G_A \left( Q^2 = 0 \right) = \lambda^\mathrm{exp} g_V - 0.0223 = 1.2850$. For the isovector-vector form factor $F_{V1}$, we multiply the tree-level result by a normalization constant $F_{V1} \left( Q^2 = 0 \right) = g_V - 0.0154 = 1.0096$.\label{footnote_FA0}}
\begin{align}
    G_A \left( \mu = M,~\xi_\gamma = 1,~Q^2 = 0 \right) &\; = \;  g_A \left(\mu_\chi = m_e \right) - 0.0223, \\
    F_{V1} \left( \mu = M,~\xi_\gamma = 1,~Q^2 = 0 \right) &\; = \;  g_V \left(\mu_\chi = m_e \right) - 0.0154.
\end{align}
For the axial-vector radius $r_A$, we exploit the conventional phenomenological definition:
\begin{align}\label{eq:rAdef}
    r_A^2  &\; = \;  {- \frac{6}{G_A \left( 0 \right)} \frac{\mathrm{d} G_A \left( Q^2 \right)}{ \mathrm{d} Q^2} \Bigg |_{Q^2 = 0}}.
\end{align}

\section{$z$-expansion parameterization}
\label{sec3}

We specify the parameterization for the nucleon axial-vector form factor in the form of a $z$ expansion as~\cite{Hill:2011wy,Bhattacharya:2011ah,Meyer:2016oeg,Borah:2020gte},
\begin{align}
    G_A \left( Q^2 \right) = \sum \limits_{k=0}^{k_\mathrm{max}} a_k z \left( Q^2 \right)^k, \qquad z \left( Q^2 \right) = \frac{\sqrt{t_\mathrm{cut} + Q^2} - \sqrt{t_\mathrm{cut} - t_0}}{\sqrt{t_\mathrm{cut} + Q^2} + \sqrt{t_\mathrm{cut} - t_0}}, \label{eq:$z$-exp}
\end{align}
with the parameters $t_\mathrm{cut} = 9 m_{\pi}^2 = 0.1616~\mathrm{GeV}^2$, $t_0 = - 0.28~\mathrm{GeV}^2$ for the deuterium data~\cite{Meyer:2016oeg}, $t_0 = - 0.5~\mathrm{GeV}^2$ for the pseudodata and the hydrogen data~\cite{MINERvA:2023avz}.\footnote{The value $t_0 = - 0.75~\mathrm{GeV}^2$ that minimizes the expansion variable $z$ within the range of the available data was selected in the original analysis of the MINERvA Collaboration~\cite{MINERvA:2023avz}.} We explain the choice of the parameter $t_0$ and $k_\mathrm{max}$ in Appendix~\ref{app:t0_choice}. Five coefficients $a_k$ are expressed in terms of $k_\mathrm{max}-4$ independent parameters: perturbative QCD behavior at large squared momentum transfer $Q^2$~\cite{Lepage:1980fj,Chernyak:1983ej} implies four sum rules~\cite{Becher:2005bg,Lee:2015jqa,Meyer:2016oeg}:
\begin{align}
    \sum \limits_{k=n}^{k_\mathrm{max}} k \left( k - 1 \right) ... \left( k - n + 1 \right)  a_k = 0, \qquad n = 0, 1, 2, 3,
\end{align}
and the normalization is fixed as a form factor value at $Q^2 = 0$, i.e., $G_A \left( 0 \right) = \sum \limits_{k=0}^{k_\mathrm{max}} a_k z \left( 0 \right)^k$, with numerical values $G_A \left( 0 \right) = 1.2754$ for tree-level fits and $G_A \left( 0 \right) = 1.2850$ for fits with the radiative corrections. Following the common practice of Refs.~\cite{Meyer:2016oeg,Borah:2020gte}, we truncate the $z$ expansion of the nucleon axial-vector form factor after $k_\mathrm{max} = 8$ and vary the $4$ free parameters $a_1,~a_2,~a_3,~a_4$. In many cases, $k_\mathrm{max} = 6$ is a sufficient description of the form factor and uncertainty, and we also perform fits with $k_\mathrm{max} = 7$. For all values of $a_k$, $k = 0 \dots k_\mathrm{max}$, we add a penalty term to $\chi^2$ with Gaussian prior, since the coefficients $a_k$ are bounded and must decrease in size for sufficiently large $k$~\cite{Hill:2010yb,Bhattacharya:2011ah}, by assigning the uncertainty $\delta a_k = 10^{-\frac{\lambda}{2}} \times |a_0| \times {\rm min}\Big(5, \frac{25}{k}\Big)$, with the regularization parameter $\lambda$ and its default value $\lambda = - 1$, cf. Appendix~\ref{app:t0_choice} for the discussion of this choice. For this default value, the size of the bound $a_0$ on all coefficients of $z$ expansion is estimated at unity order in the dispersive analysis of Refs.~\cite{Hill:2010yb,Bhattacharya:2011ah}.

\section{Radiative corrections}
\label{sec4}

For the extraction of process-independent nucleon structure in this work, we apply the QED radiative corrections to predictions of the experimental measurements, leaving the data itself untouched. In this Section, we specify the type of radiative corrections for the application to the experimental data and simulations in this work.

Since the radiated photon was not detected in both deuterium and hydrogen experimental measurements, data points effectively represent an inclusive measurement w.r.t. the kinematics of the photon. We take the fixed-order QED calculation for charged-current elastic (anti)neutrino-nucleon scattering from Refs.~\cite{Tomalak:2021hec,Tomalak:2022xup}, without including the inelastic excitations~\cite{Tomalak:2025vxa}. For tree-level diagrams, this calculation exploits the nucleon vector form factors~\cite{Borah:2020gte} and the axial-vector form factor~\cite{Meyer:2016oeg} fits in the form of $z$ expansion with resulting uncertainties at the level of cross-section ratios at the permille size and below. For modeling the hadron physics in loop diagrams, we use the dipole form for the nucleon form factors and find the corresponding uncertainties due to the form-factor choice below one permille. We assume that the squared momentum transfer in all experiments is reconstructed from the final-state lepton and apply one-dimensional lepton-energy-spectrum type of radiative corrections. In all of the data sets under consideration, the momentum transfer is fit to the kinematics of the visible particles, including the lepton, the struck proton, and, if it has a high enough momentum, the spectator proton. However, only the lepton energy enters the theoretical evaluation of radiative corrections, which is free from associated uncertainties in the reconstruction of the (anti)neutrino energy. From this, we predict the flux-integrated corrections for both hydrogen and deuterium scattering measurements as a function of the squared momentum transfer for a fixed set of the nucleon form factors.

\begin{figure}[hptb]
\centering
\begin{subfigure}[b]{0.49\textwidth}
    \includegraphics[width=\textwidth,trim=0mm 30mm 0mm 50mm,clip]{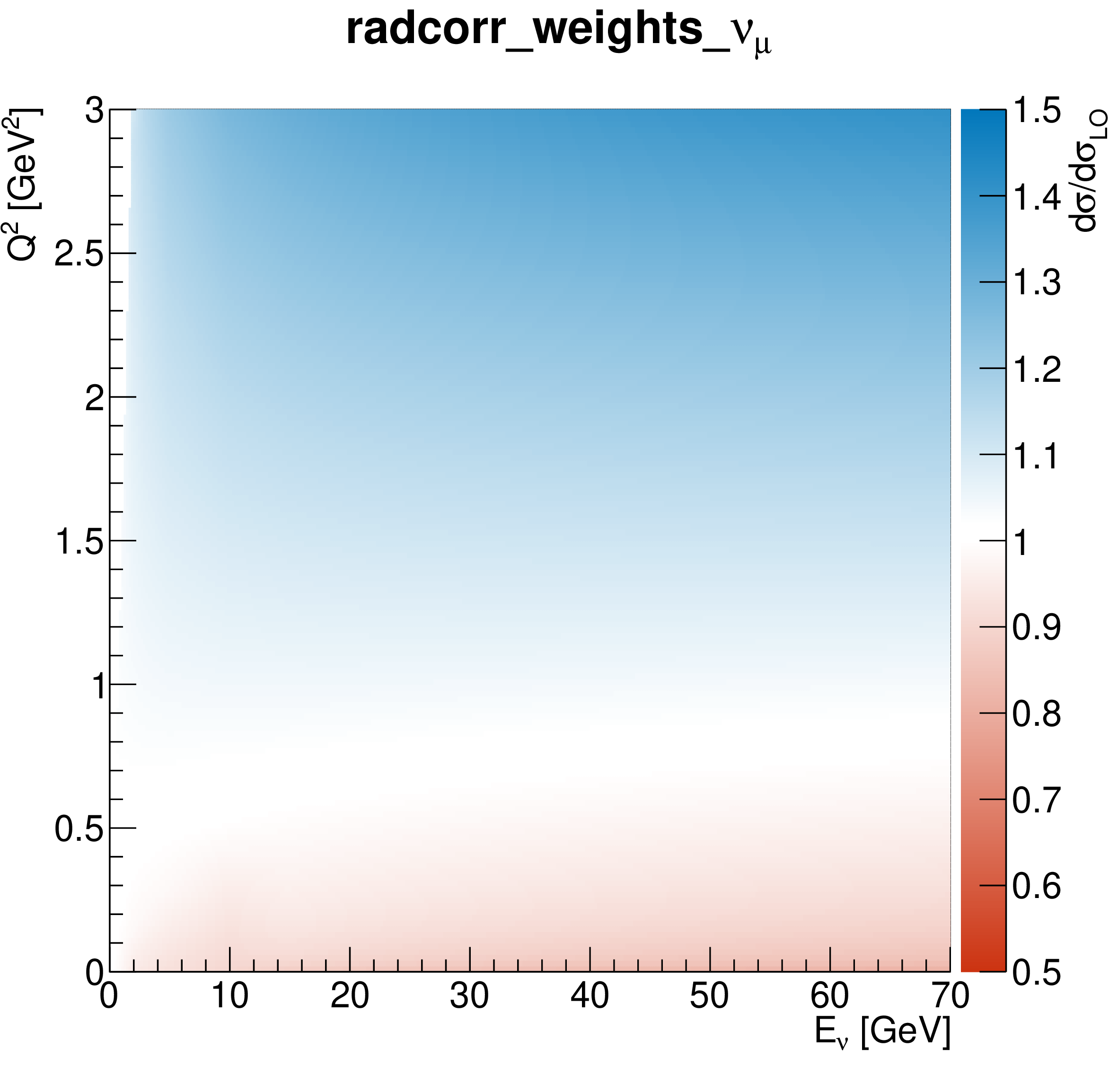}
\end{subfigure}
\begin{subfigure}[b]{0.49\textwidth}
    \includegraphics[width=\textwidth,trim=0mm 30mm 0mm 50mm,clip]{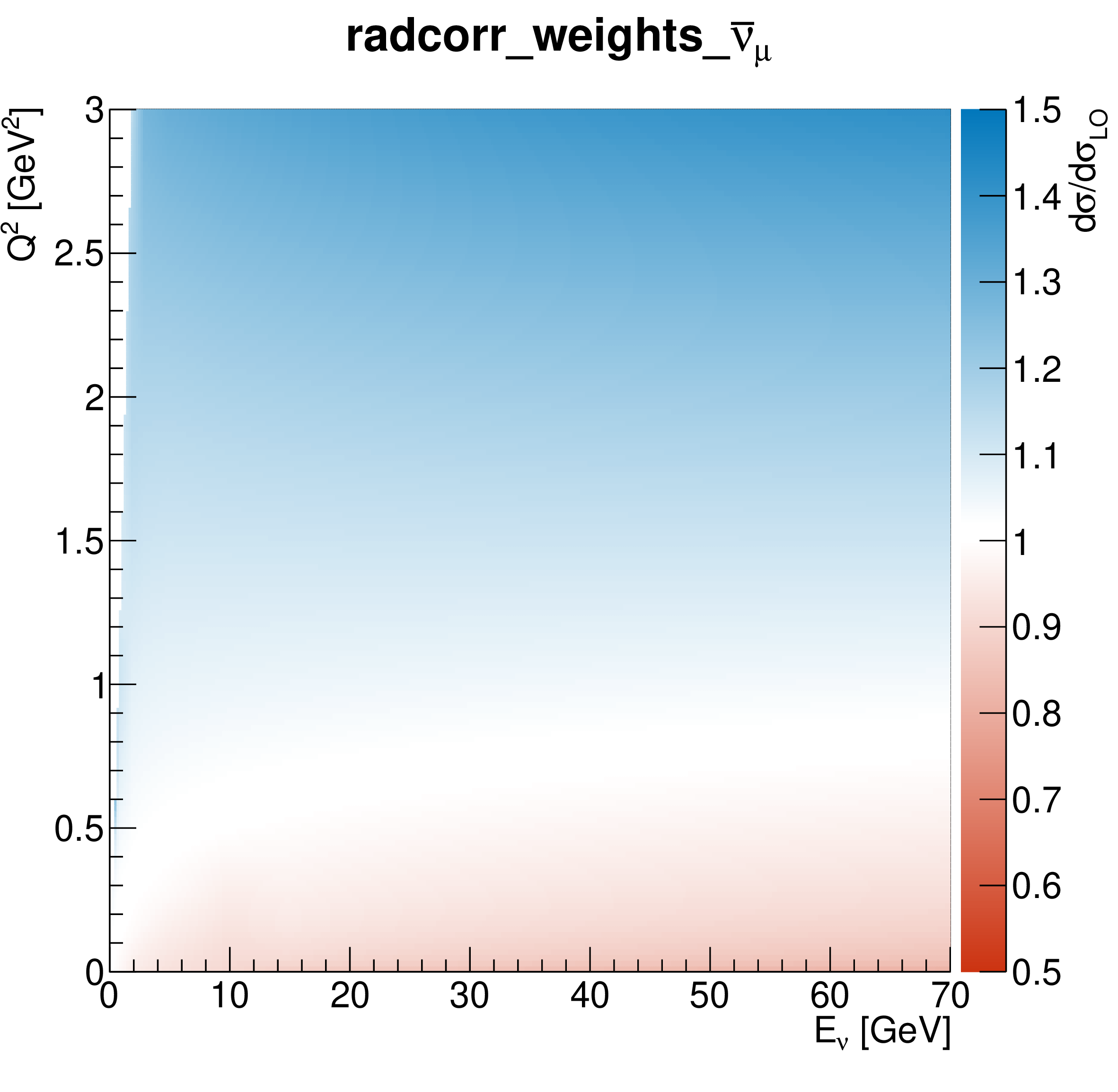}
\end{subfigure}
\caption{The ratio of the cross section with the radiative corrections to the leading-order cross section across $Q^2$ and $E_\nu$ for $\nu_\mu$ (left) and $\bar{\nu}_\mu$ (right).}
\label{fig:table_calc}
\end{figure}

\begin{figure}[hptb]
    \centering
        \includegraphics[width=0.58\textwidth]{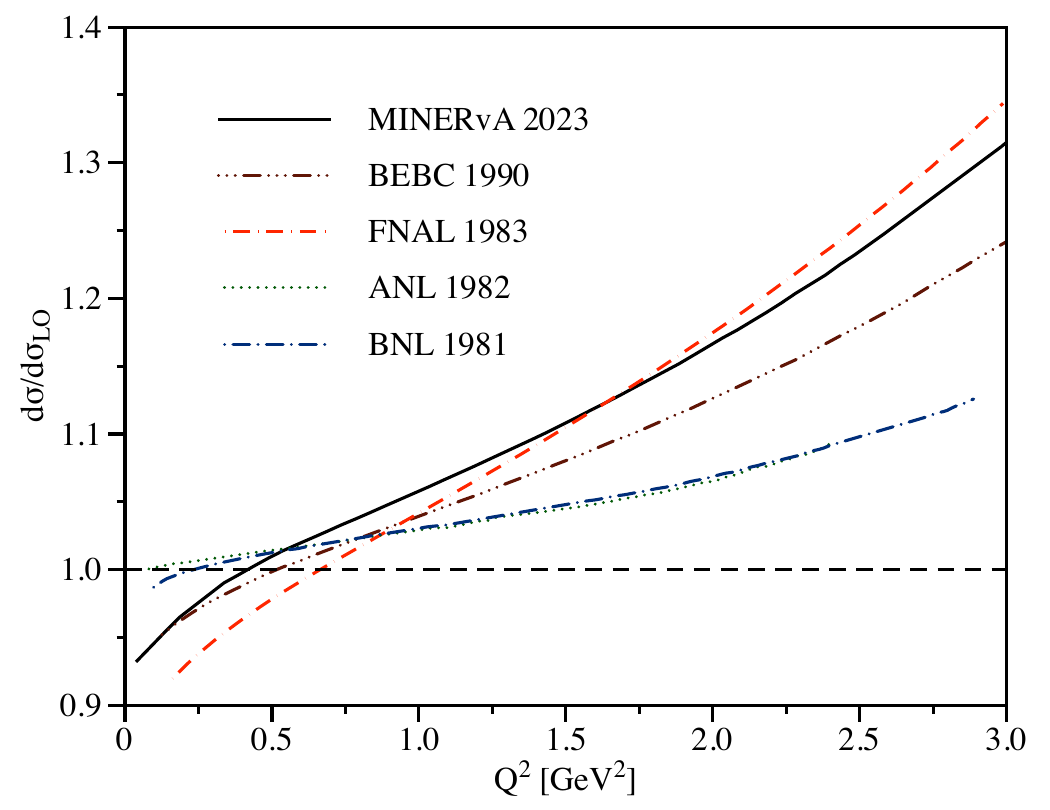}
    \hspace{0.cm}
    \caption{The ratio of the cross section after accounting for the radiative corrections relative the leading-order cross section as a function of $Q^2$, for the MINERvA (black solid line), BEBC (purple dash-dotted line), FNAL (red dashed line), ANL (green dotted line), and BNL (blue dash-dotted line) measurements. The black dashed line at $d\sigma/d\sigma_\textrm{LO}=1$ is intended to guide the eye.} 
    \label{fig:residuals_Q2}
\end{figure}

Tables are precalculated for muon (anti)neutrinos, with fixed (anti)neutrino energy $E_\nu$, and squared four-momentum transfer $Q^2$ in the struck nucleon's rest frame, as presented in Ref.~\cite{Tomalak:2022xup}. The range of (anti)neutrino energy spans $0.1$-$70~\text{GeV}$, and the $Q^2$ range covers $0$-$3~\text{GeV}^2$. A visualization of the correction in $E_\nu$ and $Q^2$ is provided in Fig.~\ref{fig:table_calc}. The radiative corrections grow with $Q^2$ for all the experiments' fluxes, as shown in Fig.~\ref{fig:residuals_Q2}, and the high-$Q^2$ region sees the largest enhancement. However, the charged-current quasielastic cross section at high $Q^2$ is very small, which limits the impact of the effect and justifies the chosen $Q^2$ range of 0-3 $\text{GeV}^2$.

To compare to an individual dataset, we evaluate the flux-averaged radiative corrections as a function of the squared momentum transfer for each set of the experimental data. The MINERvA cross sections are extracted from the scattering of muon antineutrinos on hydrogen, in contrast to other experiments which used muon neutrino beams. Furthermore, the MINERvA result is reported as a cross section over a restricted range of muon energy and momentum~\cite{MINERvA:2023avz} in the ``medium energy'' flux configuration. The ratio of the cross section after accounting for the radiative corrections to the leading-order cross section is shown in Fig.~\ref{fig:residuals_Q2} for the MINERvA, ANL, BEBC, BNL, and FNAL muon (anti)neutrino data as a function of the squared four-momentum transfer. Hence the radiative corrections are enhanced by a single logarithm of the hard energy scale over the muon mass and important for the analysis of the FNAL and BEBC neutrino and MINERvA antineutrino data and are less important in ANL and BNL neutrino measurements, due to the higher beam energy in the experiments at Fermilab. Similarly, the impact is small for the T2K and Hyper-K experiments, as discussed in Ref.~\cite{Tomalak:2022xup}.

\section{Application to (anti)neutrino scattering data}
\label{sec5}

In this Section, we present updated fits to hydrogen data for various choices of the nucleon vector form factors, with and without applying the radiative corrections. We provide $z$-expansion parameters, quantify the quality of the fit, and present the corresponding squared nucleon axial-vector radius, $r_A^2$, in Section~\ref{sec:hydrogen_data}. Covariance matrices and values for $z$-expansion parameters with larger amount of significant digits are presented in the Supplementary Material for fits to the experimental data. Having an erroneously strong dependence of fits to the deuterium data on the regularization, as illustrated in Appendix~\ref{app:t0_choice}, we provide studies of effects from radiative corrections on the pseudodata for the DUNE, Hyper-K, MINERvA, and BEBC fluxes in Section~\ref{sec:deuterium_simulation}.

\subsection{MINERvA data} 
\label{sec:hydrogen_data}

In this Section, we investigate the impacts of radiative corrections and various choices of the nucleon vector form factors on the extraction of the nucleon axial-vector form factor and radius from the MINERvA hydrogen data. The systematic uncertainties at low $Q^2$ have been extensively evaluated for the MINERvA data, so no cut to the low-$Q^{2}$ region is applied, unlike the deuterium data.

\begin{figure}
    \centering
\begin{subfigure}[b]{0.49\textwidth}
    \includegraphics[width=\textwidth,trim=0mm 0mm 16mm 8mm,clip,page=1]{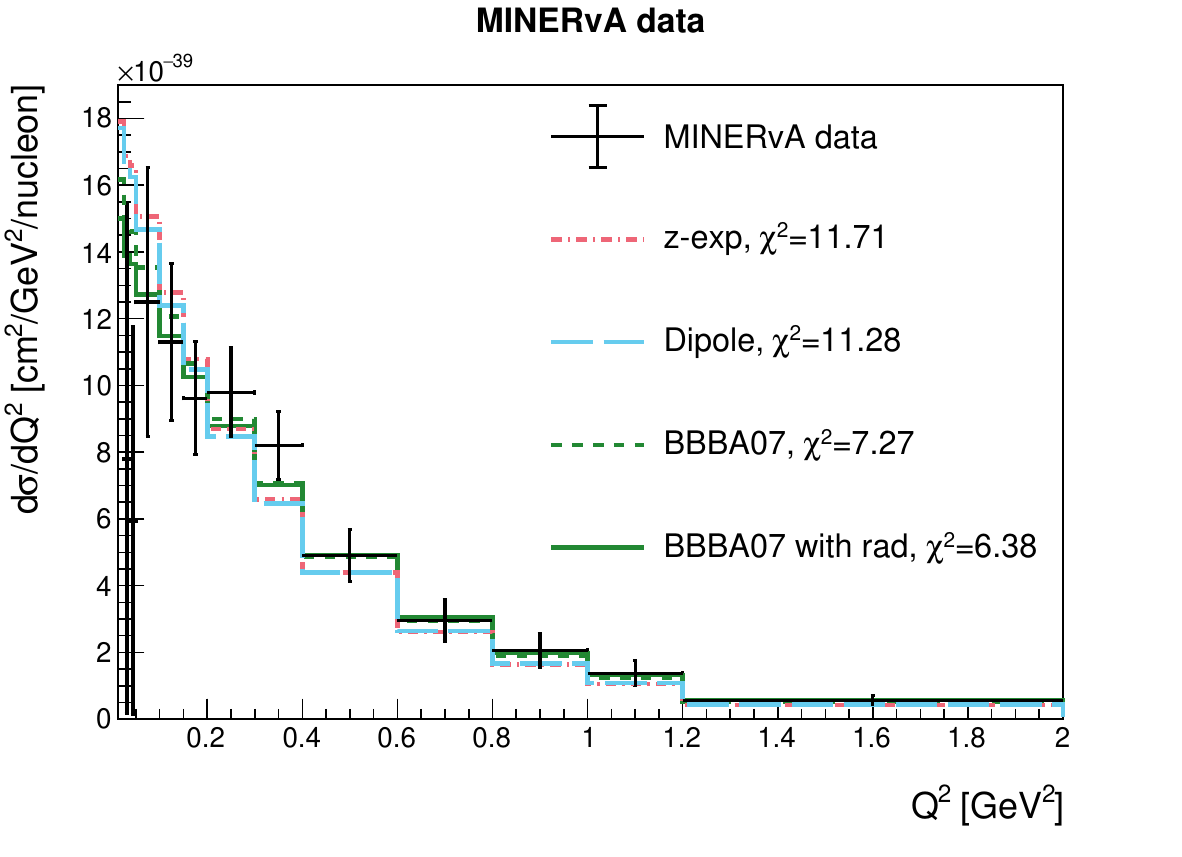}
\end{subfigure}
\begin{subfigure}[b]{0.49\textwidth}
    \includegraphics[width=\textwidth,trim=0mm 0mm 16mm 8mm,clip,page=2]{Figures/radcorr_axff_comp.pdf}
\end{subfigure}
\caption{MINERvA muon antineutrino-hydrogen charged-current quasielastic data are compared to predictions with three different nucleon axial-vector form-factor choices from Refs.~\cite{Bodek:2007vi,Meyer:2016oeg}, where the BBBA07 form-factor choice is taken with and without the radiative corrections applied. The left figure shows the differential cross section in squared four-momentum transfer, and the right shows the data and calculations relative to the BBBA07 prediction without the radiative corrections. The calculated $\chi^2$ utilizes the covariance matrix provided by the MINERvA Collaboration.} \label{fig:MINERvA_H_models}
\end{figure}

Figure~\ref{fig:MINERvA_H_models} shows the predictions for three different nucleon axial-vector form-factor choices~\cite{Meyer:2016oeg,Bodek:2007ym}, with the vector form factors taken from Ref.~\cite{Bradford:2006yz}, against the MINERvA data. Comparing the prediction based on the axial-vector form factor BBBA07~\cite{Bodek:2007vi} to the dipole form with the axial-vector mass $M_A = 1.014~\mathrm{GeV}$ and $z$-expansion parameterizations~\cite{Meyer:2016oeg}, the latter predicts a larger cross section at lower $Q^2$, and the radiative corrections further suppress the low-$Q^2$ cross section, and enhance the cross section above $0.5~\text{GeV}/c^2$. Although the $\chi^2$ for all parameterizations with vs without radiative corrections do not change by more than $1$, the radiative corrections generally reduce $\chi^2$ and improve the agreement with data for all form-factor choices. Including the radiative corrections for this data set is a similar sized effect to changing the parameterization of the form factor. For the BBBA07 parameterization with radiative corrections, the resulting $\chi^2$ is the smallest one and cross sections follow the data extremely well.

\begin{figure*}[hptb]
    \centering
    \includegraphics[width=0.58\textwidth]{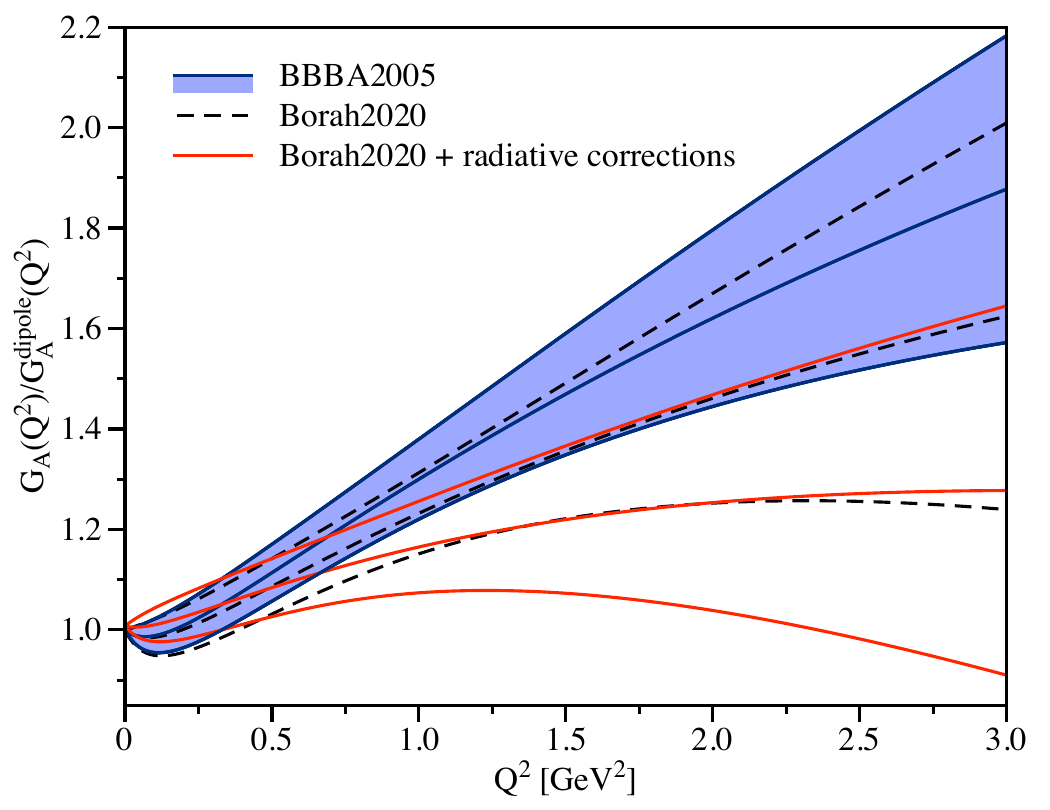}
    \hspace{0.cm}
    \caption{The nucleon axial-vector form factor $G_A \left( Q^2 \right)$ ratio to the dipole form $G^\mathrm{dipole}_A \left( Q^2 \right) $ with the axial-vector mass $M_A = 1.014~\mathrm{GeV}$ and the ``leading-order" $g_A = 1.2754$ is shown as a function of the squared momentum transfer $Q^2$. The form factor is obtained from fitting recent MINERvA antineutrino-hydrogen data~\protect\cite{MINERvA:2023avz} (i) using the BBBA2005 nucleon vector form factors~\cite{Bradford:2006yz} without the radiative corrections (blue solid lines with blue error band); (ii) using the Borah2020 nucleon vector form factors~\cite{Borah:2020gte}, that account for A1@MAMI data~\cite{Bernauer:2010wm,Bernauer:2013tpr} and the proton charge radius from the muonic hydrogen~\cite{Pohl:2010zza,Antognini:2013txn}, without the radiative corrections (black dashed lines with the space between the outer two lines representing the unshaded error band); and (iii) using the Borah2020 nucleon vector form factors~\cite{Borah:2020gte} with the radiative corrections~\cite{Tomalak:2021hec,Tomalak:2022xup} (red solid lines with the space between the outer two lines representing the unshaded error band).} \label{figure:axial_form_factor_hydrogen}
\end{figure*}

\begin{figure*}[hptb]
    \centering
    \includegraphics[height=0.371\textwidth]{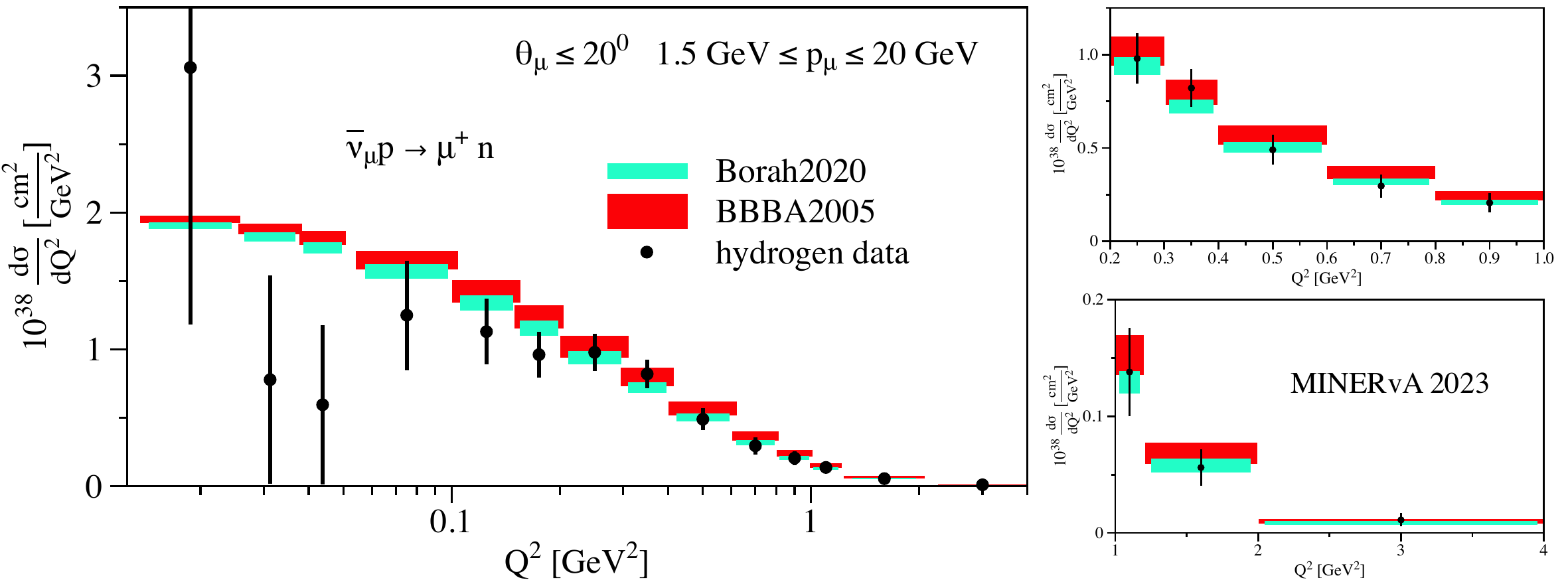}
    \caption{Antineutrino-hydrogen charged-current elastic cross-section data from MINERvA~\cite{MINERvA:2023avz} is compared with predictions based on the fits without the radiative corrections when the nucleon vector form factors are taken from Ref.~\cite{Borah:2020gte}, shown by the shorter turquoise bins, vs BBBA2005 fit~\cite{Bradford:2006yz}, shown by the red bins. Kinematic cuts in the MINERvA measurement are placed on the muon scattering angle $\theta_\mu \le 20^\circ$ and momentum $1.5~\mathrm{GeV} \le p_\mu \le 20~\mathrm{GeV}$. The thickness of the bin size in the panels represents the error, and the fifteenth bin is not shown. The two right panels zoom into the region $0.2~\mathrm{GeV}^2 \lesssim Q^2 \lesssim 1~\mathrm{GeV}^2$, and the region $Q^2 \gtrsim 1~\mathrm{GeV}^2$. Cross sections are evaluated with leading-order expressions.} \label{figure:predictions_for_MINERvA_vector_FFs}
\end{figure*}

\begin{figure*}[hptb]
    \centering
    \includegraphics[height=0.371\textwidth]{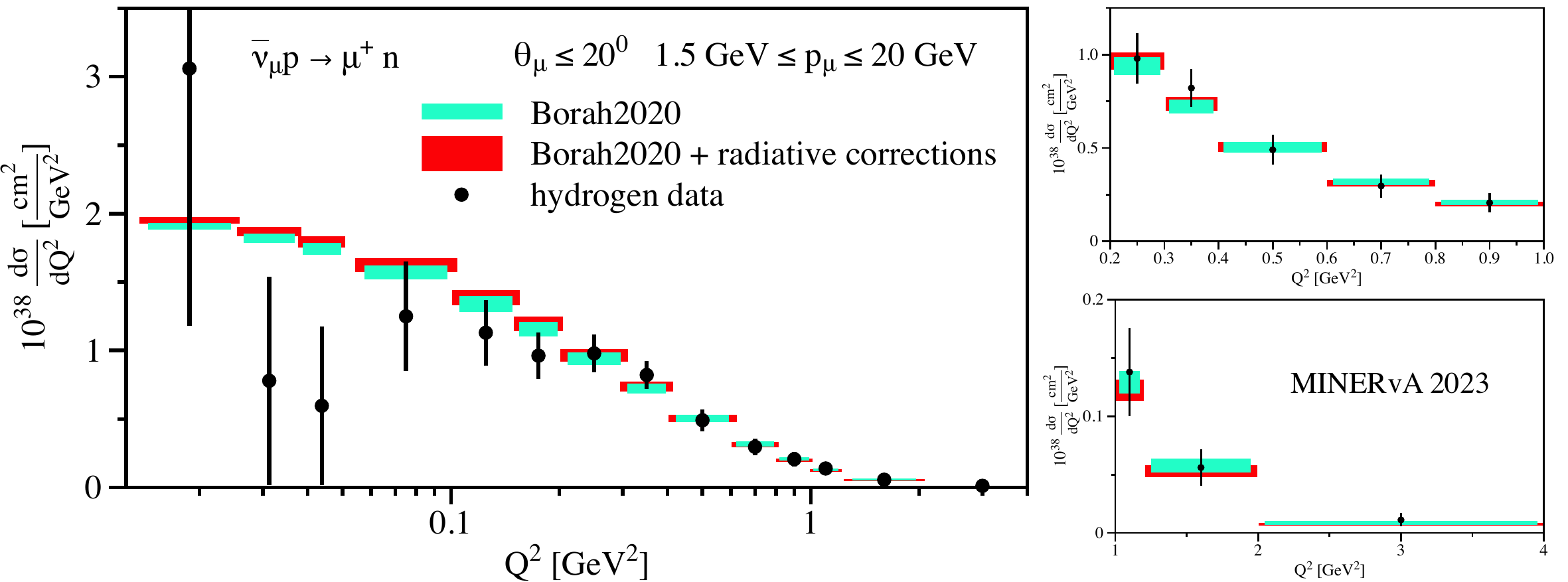}
    \caption{Same as Fig.~\ref{figure:predictions_for_MINERvA_vector_FFs} but with comparison of the fit accounting for the radiative corrections~\cite{Tomalak:2021hec,Tomalak:2022xup} to the fit without the radiative corrections, when the nucleon vector form factors are fixed to the Borah2020 parameterization from Ref.~\cite{Borah:2020gte}.} \label{figure:predictions_for_MINERvA_rad_corr}
\end{figure*}

In Table~\ref{tab:hydrogenFits_MINERvA}, we present fit results for the nucleon axial-vector form factor and radius from the antineutrino-hydrogen measurement from MINERvA, with $k_\mathrm{max} = 8$, $t_0 = - 0.5~\mathrm{GeV}^2$, and $\lambda = -1$, as discussed in Appendix~\ref{app:t0_choice}, for both BBBA2005 nucleon vector form factors~\cite{Bradford:2006yz} and Borah2020 $z$-expansion fit for the nucleon vector form factors~\cite{Borah:2020gte}, which accounts for A1@MAMI data~\cite{Bernauer:2010wm,Bernauer:2013tpr}. Fits with and without the radiative corrections applied to the cross sections are shown. In Fig.~\ref{figure:axial_form_factor_hydrogen}, we present the resulting nucleon axial-vector form factor as a function of the squared momentum transfer. To illustrate the effect of changes to the vector form factors after accounting for A1@MAMI data~\cite{Bernauer:2010wm,Bernauer:2013tpr} and the proton charge radius from the muonic hydrogen~\cite{Pohl:2010zza,Antognini:2013txn}, we provide results for both parameterizations of the vector form factors but without including the radiative corrections in the analysis. To illustrate the effect of radiative corrections, we present the results that are based on the Borah2020 $z$-expansion fit for the nucleon vector form factors~\cite{Borah:2020gte}, including the radiative corrections in the analysis. We also compare these directly to the measurement by MINERvA in Figs.~\ref{figure:predictions_for_MINERvA_vector_FFs} and~\ref{figure:predictions_for_MINERvA_rad_corr}. Overall, accounting for the radiative corrections significantly improves the description of the experimental data, while the fits illustrate a mild dependence on the nucleon vector form factors.
\begin{table}[hptb]
    \centering
    \begin{tabular}{l|c|c|c|c|c|c}
 Vector form factors and QED & $\chi^2$ & $a_1$ &  $a_2$ &  $a_3$ &  $a_4$ & $r^2_A,~\mathrm{fm}^2$\\ \hline
\text{BBBA2005}                              &  8.64 & $-1.65\pm0.15$ & $0.6\pm0.4$ & $0.7\pm1.1$ & $1.6\pm2.7$ & $0.57\pm0.15$ \\
\text{BBBA2005+radiative corrections}        &  7.78 & $-1.80\pm0.15$ & $0.7\pm0.4$ & $1.5\pm1.1$ & $0.3\pm2.5$ & $0.49\pm0.14$ \\
\text{Borah2020}                     &  9.05 & $-1.71\pm0.17$ & $0.6\pm0.4$ & $1.0\pm1.2$ & $1.5\pm2.7$ & $0.57\pm0.17$ \\
\text{Borah2020+radiative corrections} &  8.16 & $-1.88\pm0.17$ & $0.7\pm0.4$ & $2.0\pm1.2$ & $0.2\pm2.6$ & $0.49\pm0.14$ \\
    \end{tabular}
    \caption{$z$-expansion fit parameters for the nucleon axial-vector form factor $G_A \left( Q^2 \right)$, with $k_\mathrm{max} = 8$, $t_0 = - 0.5~\mathrm{GeV}^2$, and $\lambda = -1$, for the analysis of $15$ data points of the muon antineutrino-hydrogen data from MINERvA, with and without the radiative corrections. The nucleon vector form factors are taken either from BBBA2005 fit of Ref.~\cite{Bradford:2006yz} or from Borah2020 fit of Ref.~\cite{Borah:2020gte}.} \label{tab:hydrogenFits_MINERvA}
\end{table}

Following the recommendations from the detailed analysis of the elementary target data~\cite{Meyer:2025rzh}, we also present results for the nucleon axial-vector form factor and radius from the MINERvA antineutrino-hydrogen measurement, with $k_\mathrm{max} = 6$, $t_0 = - 0.5~\mathrm{GeV}^2$, unregularized, as discussed in Appendix~\ref{app:t0_choice}. These are shown in Table~\ref{tab:hydrogenFits_MINERvA_kmax6_unreg}.
\begin{table}[hptb]
    \centering
    \begin{tabular}{l|c|c|c|c}
Vector form factors and QED & $\chi^2$ & $a_1$ &  $a_2$  & $r^2_A,~\mathrm{fm}^2$\\ \hline
\text{BBBA2005}                              &  8.63 & $-1.60\pm0.22$ & $0.8\pm0.3$ & $0.58\pm0.21$ \\
\text{BBBA2005+radiative corrections}        &  7.87 & $-1.79\pm0.21$ & $1.0\pm0.3$ & $0.45\pm0.20$ \\
\text{Borah2020}                     &  9.16 & $-1.65\pm0.24$ & $0.9\pm0.3$ & $0.58\pm0.22$ \\
\text{Borah2020+radiative corrections} &  8.43 & $-1.86\pm0.23$ & $1.2\pm0.3$ & $0.43\pm0.21$ \\
    \end{tabular}
    \caption{Same as Table~\ref{tab:hydrogenFits_MINERvA} but for the fit with $k_\mathrm{max} = 6$, $t_0 = - 0.5~\mathrm{GeV}^2$, unregularized.} \label{tab:hydrogenFits_MINERvA_kmax6_unreg}
\end{table}

\subsection{DUNE, Hyper-K, MINERvA, and BEBC pseudodata}
\label{sec:deuterium_simulation}

In this Section, we study the effects of radiative corrections on the nucleon axial-vector form factor and radius that are extracted from simulated pseudodata. These pseudodata samples were created assuming muon neutrino fluxes from the DUNE~\cite{DUNE:2015lol,DUNE:2020ypp} and Hyper-K~\cite{Hyper-Kamiokande:2018ofw} technical design reports, and for the fluxes from the MINERvA~\cite{MINERvA:2022vmb,MINERvA:2023avz} and BEBC~\cite{Wachsmuth:1028107,Allasia:1990uy} experiments. The pseudodata samples are generated assuming a $1\%$ statistical uncertainty in the first bin. The other bins are scaled by their central values to mimic $\sqrt{N}$ scaling relative to the first bin.

We generate ten random flux-integrated differential cross sections for four choices of the incoming muon neutrino flux at DUNE, Hyper-K, MINERvA, and BEBC experiments, to act as ``near detectors'' in oscillation experiments. We use 51 $Q^{2}$ bins from $0$ to $2.04~{\rm GeV}^{2}$ equally spaced in increments of $0.04~{\rm GeV}^{2}$. We generate the pseudodata using the nucleon vector form factors of Ref.~\cite{Borah:2020gte} and the axial-vector form factor from the hydrogen data of Ref.~\cite{MINERvA:2023avz} separately with and without the radiative corrections. The results are averaged over the randomly generated ten datasets. To study the effects from radiative corrections, we perform fits of the nucleon axial-vector form factor for the fixed vector form factors from Ref.~\cite{Borah:2020gte} with or without the radiative corrections applied to the fit model, as it is discussed in Section~\ref{sec4}, and also vary the number of $z$-expansion coefficients between $k_\mathrm{max} = 6$, without regularization with $53$ degrees of freedom, and $k_\mathrm{max} = 7$ with a regularization $\lambda \approx - 0.95$ and $51$ degrees of freedom. We specify the fit parameter $t_0 = -0.5~\mathrm{GeV}^2$.

Table~\ref{tab:pseudodatafits} presents the resulting $\chi^2$ for the fits and the squared nucleon axial-vector radius $r^2_A$, to be compared with the axial-vector radius in the simulation $\left(r^2_A\right)_\mathrm{true} = 0.57(17)~\mathrm{fm}^2$~\cite{MINERvA:2023avz}. To illustrate the effects of radiative corrections, we emphasize the difference between the cases where both the pseudodata generation and fit make the same assumptions about the radiative corrections, fit without radiative corrections (no RC) fit to the data generated without radiative corrections (no RC) or fit with radiative corrections (RC) fit to the data generated with radiative corrections (RC), versus cases where the pseudodata generation and fit make different assumptions, no RC fit to RC data or RC fit to no RC data. We present such differences at the level of the nucleon axial-vector form factor in Figs.~\ref{figure:pseudodata}. The dependence on the fit parameter $t_0$ is explored in Appendix~\ref{app:t0_variation}.

The uncertainties reported in Tables~\ref{tab:pseudodatafits}, ~\ref{tab:pseudodatafits2}, and~\ref{tab:pseudodatafits3} in Appendix~\ref{app:t0_variation} are a consequence of the assumed statistical uncertainty of the pseudodata. These uncertainties are not representative of the current precision but could indicate sensitivity to the radiative corrections for future higher-precision experiments at percent and subpercent precision level. The reported uncertainties for $k_\mathrm{max}=7$ are also larger than for $k_\mathrm{max}=6$. This can be attributed to the $z$-expansion parameterization: the extreme ends of the parameter space, i.e., at $Q^{2}=0$ and the maximum $Q^{2}$, are the most sensitive to the largest powers of $z$ in the power series, which also correspond to the least constrained $z$-expansion coefficients. The squared axial-vector radius, which is computed at $Q^{2}=0$, is therefore sensitive to this choice. For the input form factors, we use the model with $k_\mathrm{max}=6$ or $7$ parameterization. It would not make sense to explore higher-order fit parameterizations without increasing the order of the input form factors as well.

\begin{table}[hptb]
    \centering
\begin{tabular}{l|c|c|c|c}
        &\multicolumn{2}{c|}{$k_\mathrm{max} = 6$} &\multicolumn{2}{c}{$k_\mathrm{max} = 7$} \\  \cline{2-5}  & & & & \\[-2.5ex]
 Flux mode and QED &  $\chi^2$ & $\left(r_A^2\right)_\mathrm{true} - r^2_A,~\mathrm{fm}^2$ &   $\chi^2$ & $\left(r_A^2\right)_\mathrm{true} - r^2_A,~\mathrm{fm}^2$ \\[0.4ex] \hline \hline
DUNE              &  &  & \\ \hline
\text{RC} fit to \text{RC} data &  53.3 &  0.064(12) &  50.0 &  0.114(22) \\
\text{RC} fit to \text{no RC} data &  52.0 &  -0.035(12) &  51.4 &  -0.031(23) \\
\text{no RC} fit to \text{RC} data &  61.6 &  0.145(14) &  59.7 &  0.175(26) \\
\text{no RC} fit to \text{no RC} data &  43.7 &  0.038(12) &  42.7 &  0.033(22) \\
\hline
Hyper-K           &  &  & \\ \hline
\text{RC} fit to \text{RC} data &  56.5 &  0.075(20) &  65.2 &  0.071(34) \\
\text{RC} fit to \text{no RC} data &  61.2 &  0.098(20) &  57.0 &  0.165(31) \\
\text{no RC} fit to \text{RC} data &  50.9 &  0.007(20) &  49.7 &  -0.029(30) \\
\text{no RC} fit to \text{no RC} data &  49.3 &  0.028(19) &  49.1 &  0.029(30) \\
\hline
MINERvA           &  &  & \\ \hline
\text{RC} fit to \text{RC} data &  55.8 &  0.056(12) &  54.0 &  0.084(23) \\
\text{RC} fit to \text{no RC} data &  70.4 &  -0.098(13) &  66.6 &  -0.147(25) \\
\text{no RC} fit to \text{RC} data &  80.8 &  0.190(15) &  75.1 &  0.253(28) \\
\text{no RC} fit to \text{no RC} data &  48.5 &  0.032(28) &  47.1 &  0.024(22) \\
\hline
BEBC              &  &  & \\ \hline
\text{RC} fit to \text{RC} data &  98.2 &  0.232(16) & 92.8 &  0.295(31) \\
\text{RC} fit to \text{no RC} data &  46.5 &  0.032(37) & 45.5 &  0.031(21) \\
\text{no RC} fit to \text{RC} data &  94.5 &  0.228(16) & 90.2 &  0.280(30) \\
\text{no RC} fit to \text{no RC} data &  47.3 &  0.029(20) & 45.8 &  0.016(21) \\
\hline
\end{tabular}
    \caption{The difference between the true value of the squared nucleon axial-vector radius $\left(r^2_A\right)_\mathrm{true}$ and the fit result $r_A^2$, and $\chi^2$ of the fit are presented for fits to the DUNE, Hyper-K, MINERvA, and BEBC pseudodata. The number of $z$-expansion coefficients is taken as $k_\mathrm{max} = 6$, without regularization, or $k_\mathrm{max} = 7$, with a regularization $\lambda \approx -0.954$. Generated pdeudodata and fit model include or do not include the radiative corrections. The nucleon vector form factors are taken from the Borah2020 $z$-expansion fit of Ref.~\cite{Borah:2020gte}. The $z$-expansion parameter $t_0 = -0.5~\mathrm{GeV}^2$ of the nucleon axial-vector form factor is the same for generating the pseudodata and performing the fits. The axial-vector radius for generating the pseudodata is $\left(r^2_A\right)_\mathrm{true}  = 0.57(17)~\mathrm{fm}^2$~\cite{MINERvA:2023avz}.} \label{tab:pseudodatafits}
\end{table}

\begin{figure*}[hptb]
    \centering
    \includegraphics[width=1.\textwidth]{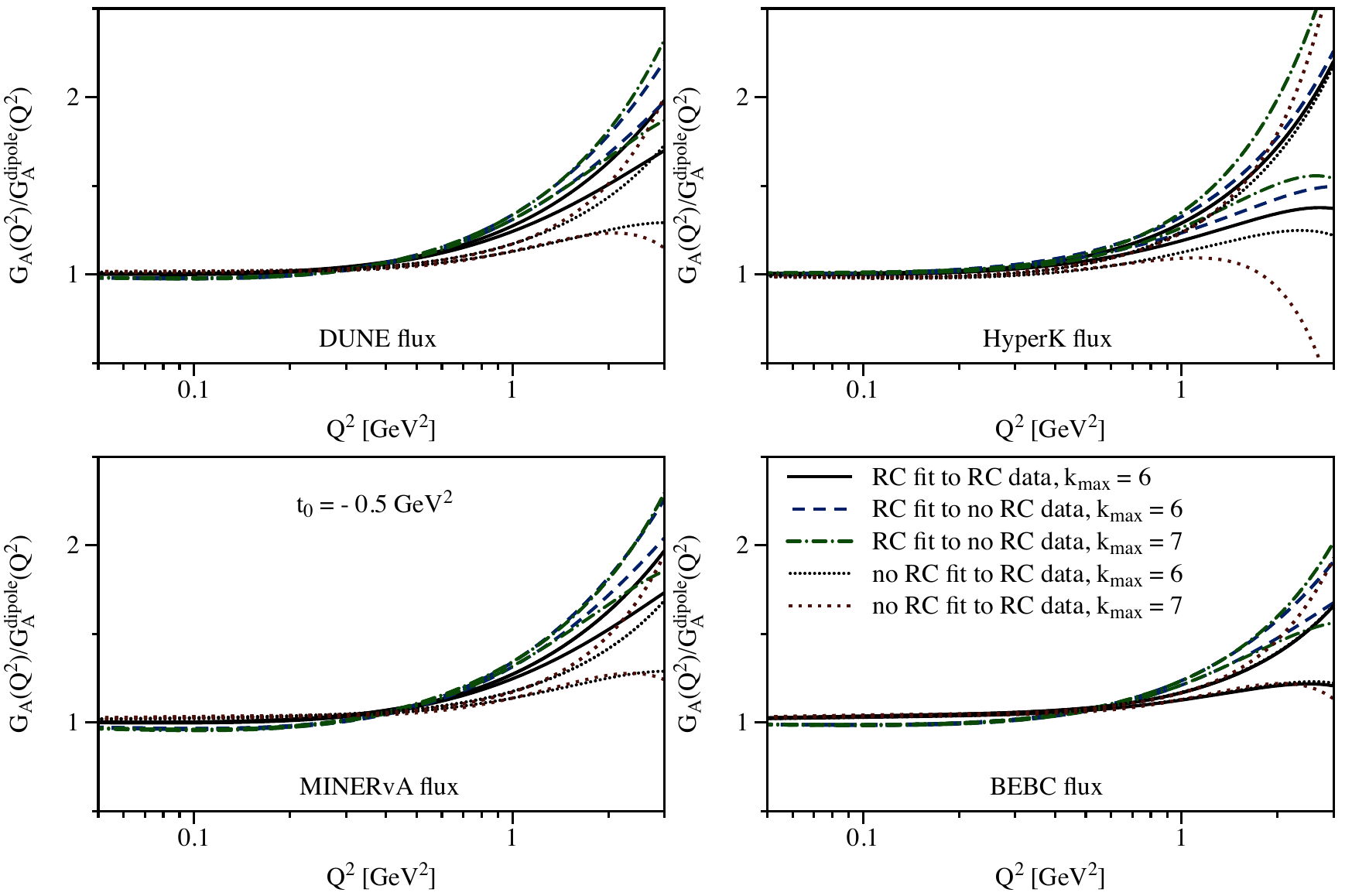}
    \hspace{0.cm}
    \caption{The nucleon axial-vector form factor $G_A \left( Q^2 \right)$ ratio to the dipole form $G^\mathrm{dipole}_A \left( Q^2 \right) $ with the axial-vector mass $M_A = 1.014~\mathrm{GeV}$ and the ``leading-order" $g_A = 1.2754$ is shown as a function of the squared momentum transfer $Q^2$ for fits to the DUNE, Hyper-K, MINERvA, and BEBC pseudodata on the upper left, upper right, lower left, and lower right panels, respectively. Results for fits and pseudodata with and without the radiative corrections are compared. Only upper and lower limits for the fits are shown.} \label{figure:pseudodata}
\end{figure*}

\section{Extrapolation to future measurements}
\label{sec6}

According to recent study~\cite{Alvarez-Ruso:2022ctb}, based on the nucleon vector form factors from Ref.~\cite{Borah:2020gte} and the axial-vector form factor from Ref.~\cite{Meyer:2016oeg}, the statistical error of the nucleon axial-vector radius extractions scale with the number of charged-current quasielastic interactions $N$ approximately as $1/\sqrt{N}$:
\begin{align}\label{eq:rA2error}
    \delta r_A &= \left(\delta r_A \right)_0 \sqrt{\frac{N_0}{N}}\,, \qquad \left(\delta r_A \right)_0 \sim 0.2-0.25~\mathrm{fm} , \qquad N_0 = 10^4 \,,
\end{align}
where the numerical values for parameters $(\delta r_A)_0$ and $N_0$ correspond to the existing ANL and BNL datasets. The number of antineutrino-hydrogen interactions is expected to be $N\sim 1.7 \times 10^{6}$~\cite{DUNE:2015lol} using $700~\mathrm{kg}$ of hydrogen in the anticipated DUNE near detector~\cite{Duyang:2019prb,Poppi:2023ufv} over an integrated exposure of $8\times10^{21}$ POT. Such exposures can improve the accuracy on the nucleon axial-vector radius to $\delta r_A = 0.015-0.020~\mathrm{fm}$, which is in reasonable agreement with recent projections~\cite{Petti:2023abz}. For the extracted nucleon axial-vector form factor, we expect a relative statistical error at the permille level.

\section{On comparisons to lattice QCD}
\label{sec7}

As of today, first-principles calculations of the nucleon axial-vector form factor with QCD on a spacetime lattice are maturing but have not yet achieved the necessary precision to probe the radiative corrections. Lattice QCD computations of the nucleon axial-vector coupling have reached agreement with experimental results at the percent level after contending with a long history of improperly controlled systematic effects~\cite{Bhattacharya:2011qm,Bar:2016uoj,Bar:2017kxh,Bar:2018xyi,Chang:2018uxx,Gupta:2018qil,RQCD:2019jai,He:2021yvm,FlavourLatticeAveragingGroupFLAG:2021npn}. These complications are likely due to contamination from the pion-nucleon excited states, created by an axial-vector current that acts as a pion creation operator. Computations of the momentum dependence of the axial-vector form factor are presently at around five-percent precision with complete error budgets~\cite{Jang:2019vkm,RQCD:2019jai,Alexandrou:2020okk,Park:2021ypf,Djukanovic:2022wru,Jang:2023zts,Alexandrou:2023qbg} with additional improvement to about two-percent precision possible under an average~\cite{Meyer:2026kdl}. Additional computations with explicit pion-nucleon interpolating operators are underway that will conclusively determine whether the excited state contamination in these calculations are fully under control. In this Section, we discuss how evaluations of the nucleon axial-vector form factor might be improved upon to reach the desired percent-level precision. We will emphasize some of the unavoidable theoretical developments that will be necessary for such a journey.

To achieve control over the (anti)neutrino-nucleon scattering cross sections and compare to lattice QCD inputs, we must formulate, evaluate, and renormalize the corresponding matrix elements and perform the matching to the experiment or hadronic model at the level of Eqs.~(\ref{eq:CCQE_amplitude})~and~(\ref{eq:CCQE_amplitudem}). The minimal setup for computing nucleon matrix elements with lattice QCD produces results with QCD contributions to all orders and no QED contributions. This choice does not correspond to either the desired invariant amplitudes nor the Born form factors. To resolve this mismatch, we must also compute the QED radiative corrections on the lattice in order to quantify the size of corrections to the axial-vector matrix elements at one-loop level in QED, including the contributions that connect the hadronic system to the charged lepton that is not explicitly simulated on the lattice~\cite{Seng:2018yzq,Ma:2023kfr,Tuo:2024bhm}. When matching the lattice QCD amplitudes to the physical point, QCD and QED renormalization effects on amplitudes can be factorized at the one-loop level but only at the expense of separate renormalization schemes for both QCD and QED. At higher orders the renormalization of QCD and QED become intertwined and it is unclear how to remove the scheme dependence.

The Euclidean time and long-distance nature of QED interactions, when confined to the finite volume of the lattice, make the realization of QED radiative corrections quite complicated on the lattice~\cite{Lucini:2015hfa,Patella:2017fgk,Davoudi:2018qpl}. The Euclidean correlation function obtained from a lattice QCD calculation with more than one current insertion is related to the desired Minkowski response by an inverse Laplace transform. Performing this inverse transform on a finite grid of points is an ill-posed problem that must be tackled with methods such as the Backus-Gilbert method, Bayesian reconstruction, or other approaches~\cite{Pijpers:1992,Hansen:2019idp,PhysRevB.96.035147,PhysRevE.95.061302,Itou:2020azb,doi:10.7566/JPSJ.89.012001,PhysRevB.101.035144,PhysRevB.98.035104,Shi:2022yqw,Chen:2021giw,Kades:2019wtd,Zhou:2023pti,Lechien:2022ieg,Wang:2021jou,10.1093/gji/ggz520,DelDebbio:2021whr,Candido:2023nnb,Horak:2021syv,Pawlowski:2022zhh,Horak:2023xfb,Bergamaschi:2023xzx,Buzzicotti:2023qdv,Huang:2023gpb,Lawrence:2024hjm,DelDebbio:2024lwm}. Since electromagnetism is a long-range force, finite volume corrections to matrix elements originating from QED effects are typically only suppressed by power-law corrections in the lattice volume, rather than exponential suppression that is more common for finite volume corrections originating from QCD. These corrections manifest as a large sensitivity to the finite extent of the lattice and various methods have been developed to address these issues~\cite{Gockeler:1989wj,Polley:1990tf,Duncan:1996xy,Hayakawa:2008an,BMW:2014pzb,Lucini:2015hfa,Endres:2015gda,Patella:2017fgk,Clark:2022wjy,Hermansson-Truedsson:2023krp}. Furthermore, Gauss's law can be satisfied on the boundaries of a finite periodic box only if the enclosed charges sum to zero~\cite{Hayakawa:2008an}, so computations must be performed either with modified boundary conditions that have nontrivial consequences to the computation symmetries or as a perturbative series in QED. Approaches for handling or circumventing these complications are an active area of research.

In addition to theoretical difficulties associated with QED on the lattice, a successful computation of QED matrix elements must overcome several practical difficulties. The correlation functions must be probed with a large number of Euclidean time separations in order to isolate the spectrum and interesting matrix elements, which results in increases the computational costs and memory usage requirements associated with carrying out the analysis. Depending on the target matrix elements, excited state contamination can also complicate the isolation of the relevant matrix elements, such as the difficulties with nucleon-pion states in the nucleon axial-vector form factor. Removal of these matrix elements will likely require the use of explicit multiparticle interpolating operators, further adding to combinatorial increases in the computational cost. In spite of all of these challenges, the first computations in this direction have been performed to address the QED radiative corrections to the neutron superallowed beta decay from a $\gamma W$ box diagram on the nucleon~\cite{Feng:2020zdc,Ma:2023kfr}.

The ``lattice QCD'' matrix element can be formulated as an expansion in the electromagnetic coupling constant in terms of the quark correlation functions, which can be computed on the lattice and include QCD contributions at all orders in the strong coupling constant. In the baryon sector, such a computation was formulated by two independent approaches~\cite{Marciano:1985pd,Cirigliano:2023fnz} for the nucleon vector coupling constant at low energies, when the chiral perturbation theory is valid. For the nucleon axial-vector coupling, the ``lattice QCD" formulation for the matrix elements was only recently completed~\cite{Cirigliano:2022hob,Cirigliano:2023fnz,Cirigliano:2024nfi}. The (anti)neutrino-nucleon scattering matrix elements have not been rigorously formulated with a proper effective field theory at moderate momentum transfers. Therefore, the momentum transfer dependence can be computed only by combining effective field theory techniques~\cite{Hill:2019xqk} with a traditional current-algebra approach for the electroweak radiative corrections~\cite{Sirlin:1977sv}.

\section{Conclusions and Outlook}
\label{sec8}

In this paper, we extracted and presented the nucleon axial-vector form factor and radius from the available muon antineutrino-hydrogen experimental data from MINERvA, applying the QED radiative corrections in such extractions for the first time. We also performed a simulation of these extractions from the pseudodata of neutrino scattering experiments with DUNE, Hyper-K, MINERvA, and BEBC fluxes as well as estimates of radiative effects in the neutrino-deuterium bubble-chamber experiments at ANL, BNL, FNAL, and BEBC. We exploited the single-differential fixed-order calculation for charged-current elastic (anti)neutrino-nucleon scattering and assumed that the squared momentum transfer is reconstructed from the final-state charged lepton energy neglecting the associated uncertainties from the reconstruction of the (anti)neutrino energy in experimental datasets. The radiative corrections do not significantly affect the extracted value of the nucleon axial-vector form factor from the ANL and BNL deuterium data due to smaller corrections and large experimental uncertainties in these experiments, but they shift the $z$-expansion parameters describing the extraction from the FNAL deuterium and MINERvA hydrogen data, which have comparable size of radiative corrections and experimental uncertainties, by up to a standard deviation. We observe the largest effect on central values from radiative corrections on the FNAL deuterium bubble-chamber data, as expected due to the higher neutrino energy. The radiative corrections do not shift the nucleon axial-vector radius outside the $1\sigma$ range of the experimental uncertainty. However, the size of these radiative corrections will clearly be important for high-precision extraction from GeV beams in future experiments. For the MINERvA measurement, the size of radiative corrections is similar in magnitude to the uncertainty of the nucleon axial-vector form factor. Future treatment of radiative corrections in high-precision experimental results would benefit from application of the corrections to measurements to the muon and proton kinematics directly.

\section*{Acknowledgments}

The work of O.T. is supported in part by the National Science Foundation of China under Grants No. 12347105, No. 12447101, and in part by the Los Alamos National Laboratory Directed Research and Development (LDRD/PRD) program under project numbers 20210968PRD4 and 20240127ER. This work is supported by the US Department of Energy through the Los Alamos National Laboratory. Los Alamos National Laboratory is operated by Triad National Security, LLC, for the National Nuclear Security Administration of U.S. Department of Energy (Contract No. 89233218CNA000001). The work of A.S.M. is supported in part by Lawrence Livermore National Security, LLC under Contract No. DE-AC52-07NA27344 with the DOE and by the Neutrino Theory Network Program Grant No. DE-AC02-07CHI11359 and DOE Award No. DE-SC0020250. The work of C.W. is supported by The Royal Society under award URF\textbackslash R1\textbackslash 241892 and the Science and Technologies Facilities Council. The work of K.S.M.\, T.C.\ and C.W. is supported in part by U.S. Department of Energy, Office of Science, Office of High Energy Physics under Award Number DE-SC0008475. The work of R.J.H. is supported by the U.S. Department of Energy, Office of Science, Office of High Energy Physics, under Award DE-SC0019095. This work was produced by Fermi Forward Discovery Group, LLC under Contract No. 89243024CSC000002 with the U.S. Department of Energy, Office of Science, Office of High Energy Physics.

This report was prepared as an account of work sponsored by an agency of the United States Government. Neither the United States Government nor any agency thereof, nor any of their employees, makes any warranty, express or implied, or assumes any legal liability or responsibility for the accuracy, completeness, or usefulness of any information, apparatus, product, or process disclosed, or represents that its use would not infringe privately owned rights. Reference herein to any specific commercial product, process, or service by trade name, trademark, manufacturer, or otherwise does not necessarily constitute or imply its endorsement, recommendation, or favoring by the United States Government or any agency thereof. The views and opinions of authors expressed herein do not necessarily state or reflect those of the United States Government or any agency thereof.

\bibliography{rA_radcorr}{}

\appendix

\section{Fit parameters for hydrogen and deuterium data} \label{app:t0_choice}

In this Appendix, we explain our choice of the fit parameters by studying $L$ curves that represent the dependence on the relative weight of $\chi^2$ from data $\chi^2_\mathrm{data}$ w.r.t. $\chi^2$ from the penalty term $\chi^2_\mathrm{penalty}$ for different values of $k_\mathrm{max}$ and $t_0$. The point at the corner of the $L$ represents the best compromise and guides the choice of the ``optimal" regularization parameters. This point minimizes the error of theoretical assumptions without having too large residuals.

The $\chi^2$ in our fits is determined by
\begin{align}
    \chi^2 &= \chi^2_\mathrm{data} \left( a_k \right) + \chi^2_\mathrm{penalty} \left( \lambda,~a_k \right), \\
    \chi^2_\mathrm{data} \left( a_k \right) &= \sum \limits_{\mathrm{bins}~i, j} \left( \sigma^\mathrm{data}_i - \sigma^\mathrm{theory}_i \left( a_k \right) \right) \left( \mathrm{cov}^{-1}\right)_{i j} \left( \sigma^\mathrm{data}_j - \sigma^\mathrm{theory}_j \left( a_k \right) \right), \\
    \chi^2_\mathrm{penalty} \left( \lambda,~a_k \right) &= 10^{\lambda} \sum \limits_{k=1}^{k_\mathrm{max}} \left(\frac{a_k}{a_0 \sigma^a_k} \right)^2,\; \qquad \sigma^a_k = {\rm min}\Big(5, \frac{25}{k}\Big).
    \label{eq:lambda_definition}
\end{align}

\begin{figure*}
    \centering
    \includegraphics[height=0.46\textwidth]{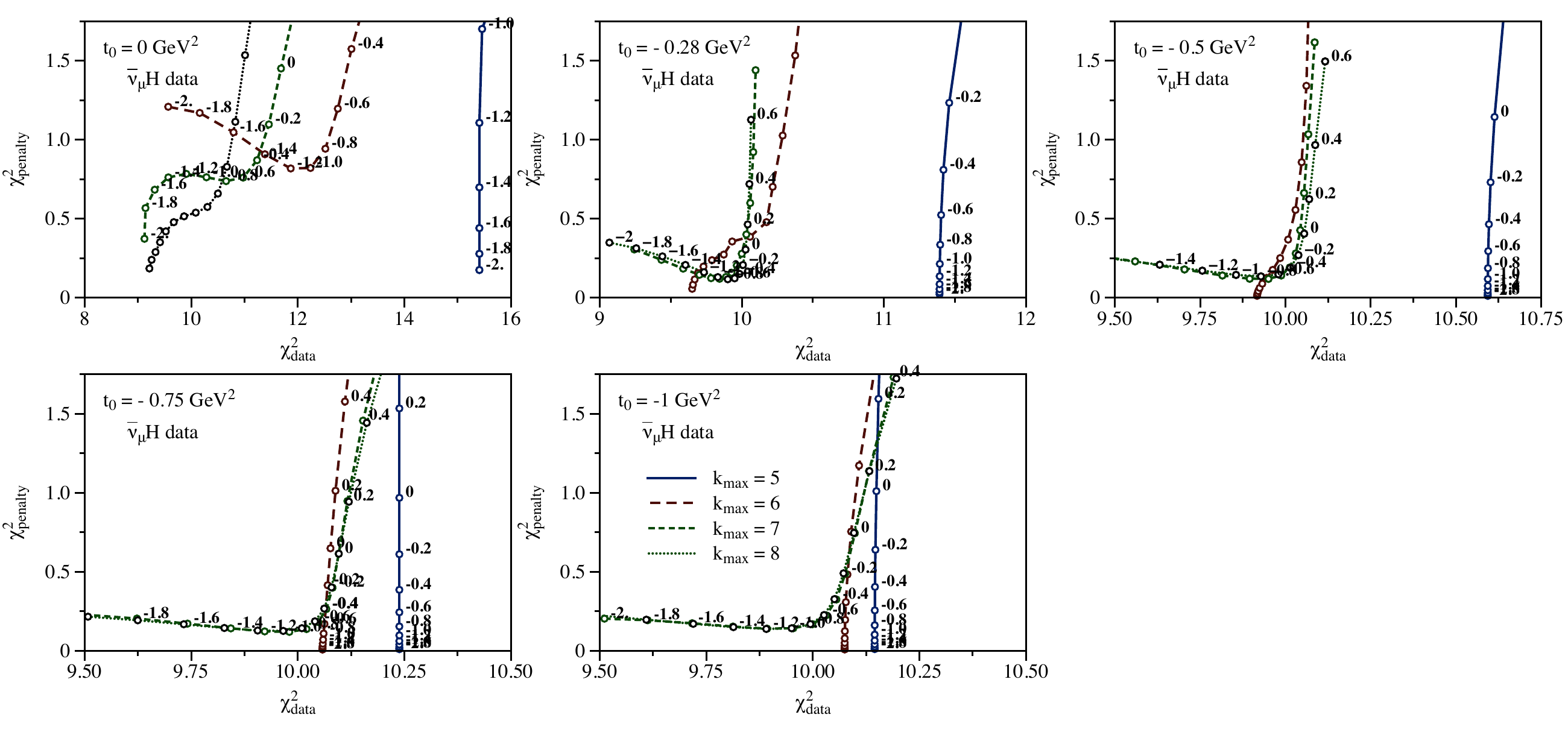}
    \caption{$L$ curves with $\chi^2_\mathrm{penalty}$ on the $y$ axis and $\chi^2_\mathrm{data}$ on the $x$ axis for the fits to the MINERvA hydrogen data with $k_\mathrm{max} = 5,~6,~7,$ and $8$ are presented for the value of the parameter $t_0 = 0,~-0.28,~-0.5,~-0.75,~-1~\mathrm{GeV}^2$, as it is indicated on figures. Curves represent the dependence on the parameter $\lambda$ that is presented for each point used in every fit as a number, cf. Eq.~(\ref{eq:lambda_definition}), and corresponds to a relative weight between $\chi^2_\mathrm{penalty}$ and $\chi^2_\mathrm{data}$ in the total $\chi^2$. Besides very low $t_0$ values, the $L$ curves for $k_\mathrm{max} = 7$ and $k_\mathrm{max} = 8$ fits are indistinguishable.} \label{fig:Lcurve_MINERvA}
\end{figure*}

First, we choose the fit parameters for the muon antineutrino-hydrogen data from the MINERvA experiment and do not include the radiative corrections in our analysis. We verified that account for the radiative corrections does not change qualitative and quantitative results in this Appendix. In Fig.~\ref{fig:Lcurve_MINERvA}, we present $L$ curves~\cite{Lcurve,Lcurvetext} for different $k_\mathrm{max} = 5,~6,~7,$ and $8$ varying the parameter $\lambda$. We indicate the value of this parameter for each point used in every fit as a number. We present our study for values of the parameter $t_0$: $t_0 = - 1~\mathrm{GeV}^2$, $t_0 = - 0.75~\mathrm{GeV}^2$, $t_0 = - 0.5~\mathrm{GeV}^2$, $t_0 = - 0.28~\mathrm{GeV}^2$, and $t_0 = 0~\mathrm{GeV}^2$. Besides very low $t_0$ values, the $L$ curves for $k_\mathrm{max} = 7$ and $k_\mathrm{max} = 8$ fits are indistinguishable. That is why we present the results in Section~\ref{sec:hydrogen_data} in a common for studies of the nucleon axial-vector form factor form with $k_\mathrm{max} = 8$~\cite{Meyer:2016oeg}. At low absolute $t_0$ values, we observe an improvement in $\chi^2$ by more than one unit for increasing $k_\mathrm{max}$ from $k_\mathrm{max} = 5$ to $k_\mathrm{max} = 6$. However, there is no improvement in $\chi^2$ for $|t_0| \gtrsim 0.5~\mathrm{GeV}^2$, and the nucleon axial-vector form factor can be described with just one free parameter. Based on $L$ curves, we can select $k_\mathrm{max} =6$ and $t_0 = - 0.5~\mathrm{GeV}^2$, when $L$ shape just starts to appear. It determines one set of our fit parameters.

\begin{figure*}
    \centering
    \includegraphics[height=0.24\textwidth]{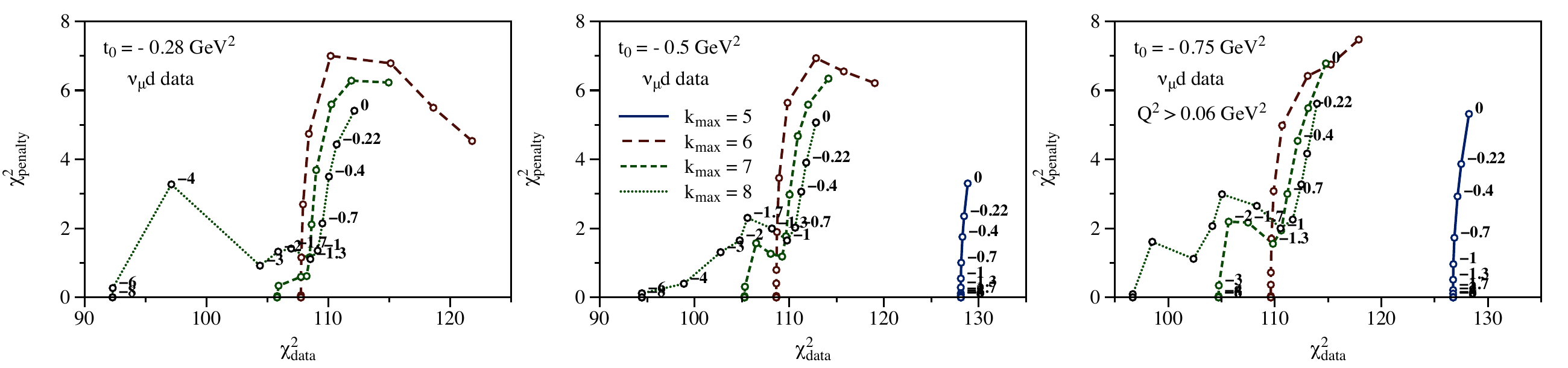}
    \caption{Same as Fig.~\ref{fig:Lcurve_MINERvA} but for the deuterium data with squared momentum transfers above $Q^2 > 0.06~\mathrm{GeV}^2$.} \label{fig:Lcurve_deuterium_006}
\end{figure*}

\begin{figure*}
    \centering
    \includegraphics[height=0.24\textwidth]{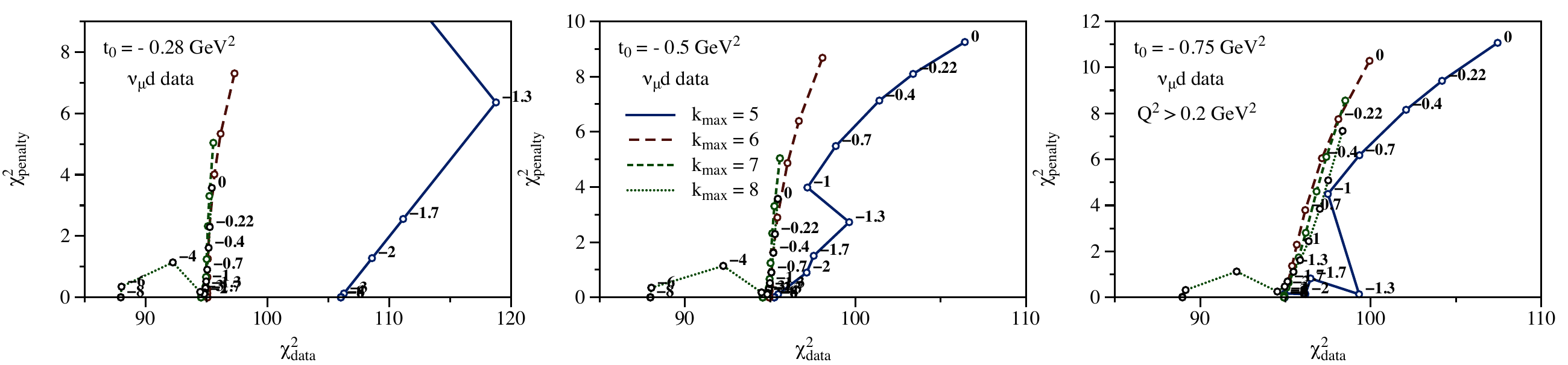}
    \caption{Same as Fig.~\ref{fig:Lcurve_MINERvA} but for the deuterium data with squared momentum transfers above $Q^2 > 0.2~\mathrm{GeV}^2$.} \label{fig:Lcurve_deuterium_02}
\end{figure*}

We repeat the same study with the deuterium bubble-chamber data of ANL, BNL, and FNAL experiments and present $L$ curves for two choices of the lowest squared momentum transfer values $Q^2_\mathrm{min} = 0.06~\mathrm{GeV}^2$ and $Q^2_\mathrm{min} = 0.2~\mathrm{GeV}^2$ in Figs.~\ref{fig:Lcurve_deuterium_006} and~\ref{fig:Lcurve_deuterium_02}, respectively. In Ref.~\cite{Meyer:2016oeg}, the deuterium bubble-chamber experiments were fit with $\lambda=0$, corresponding to a point far from the optimal bend of the $L$ curve. This gives a strong dependence on the penalty term, which when relaxed leads to a degeneracy between the axial-vector form factor parameterization and the floating normalizations used for those datasets. To avoid the complications from these difficulties, we switch from the experimental data to pseudodata for studies of effects from radiative corrections in Section~\ref{sec:deuterium_simulation}.

\newpage
\section{Fit parameters for pseudodata} \label{app:t0_variation}

\begin{figure*}[hptb]
    \centering
    \includegraphics[width=1.\textwidth]{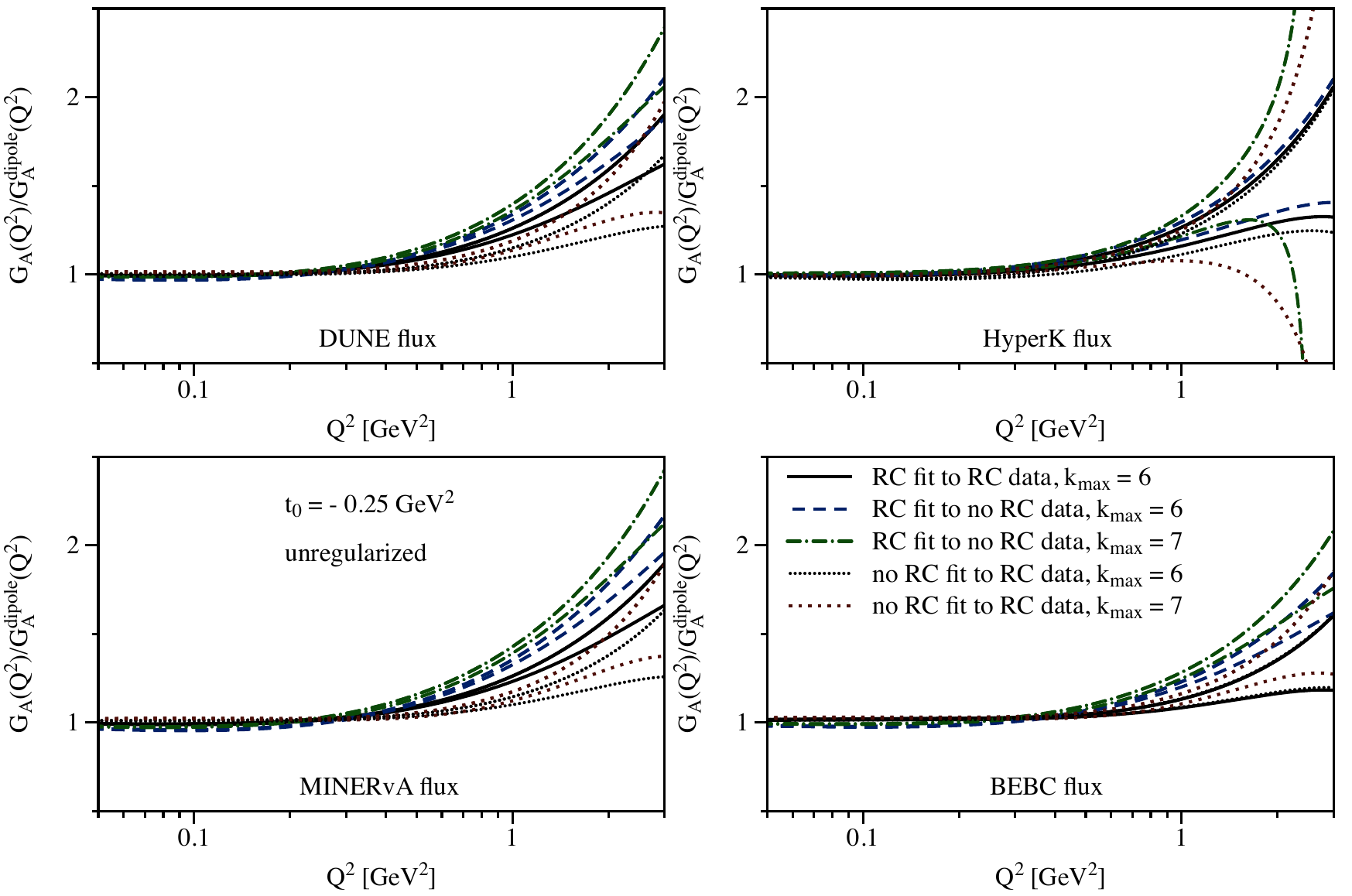}
    \hspace{0.cm}
    \caption{Same as Fig.~\ref{figure:pseudodata} but for the nucleon axial-vector form factor in the generated data from the fit to the MINERvA data with $t_0 =-0.28~\mathrm{GeV}^2$ without regularization.} \label{figure:pseudodata2}
\end{figure*}

\begin{figure*}[hptb]
    \centering
    \includegraphics[width=1.\textwidth]{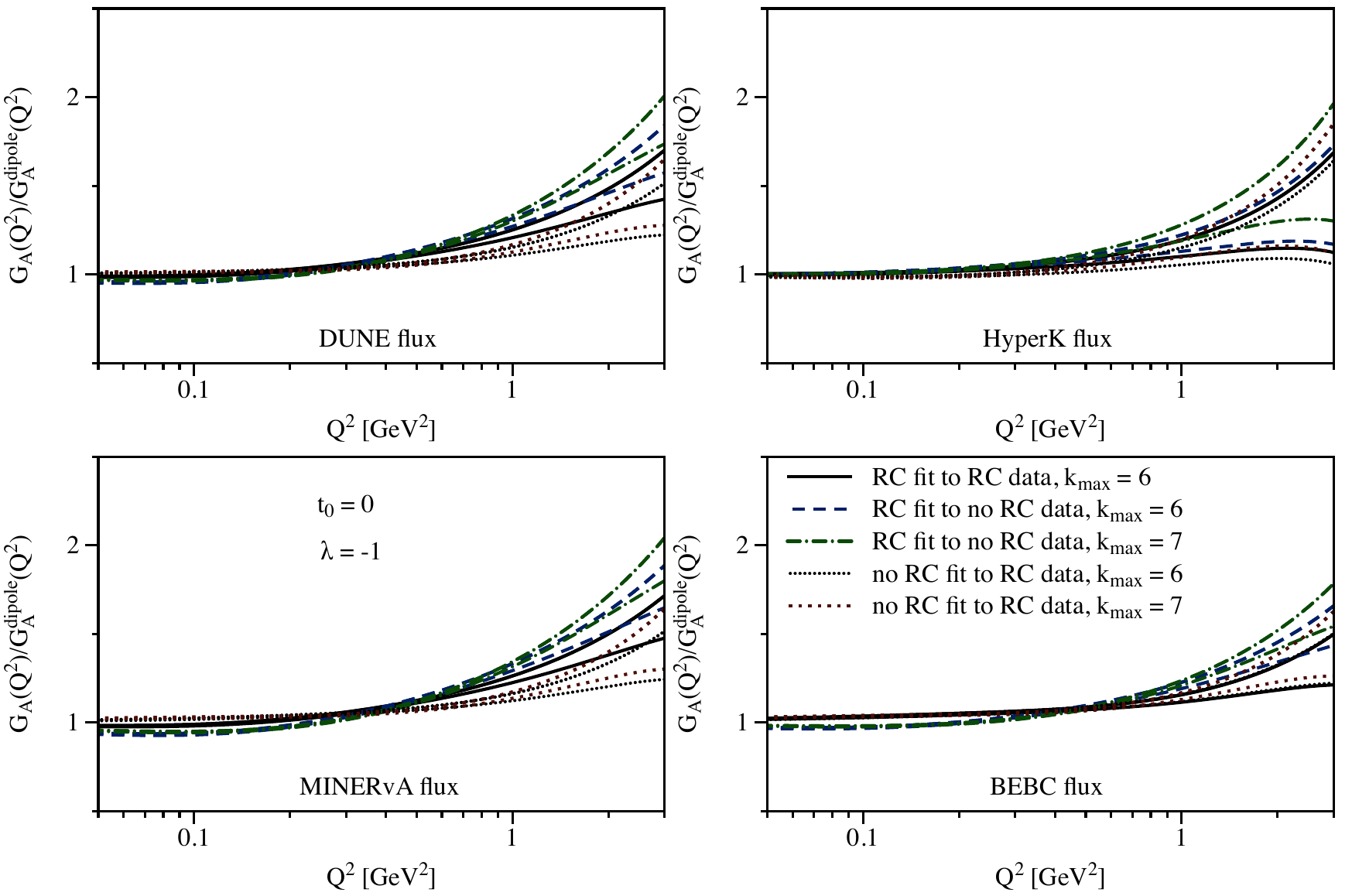}
    \hspace{0.cm}
    \caption{Same as Fig.~\ref{figure:pseudodata} but for the nucleon axial-vector form factor in the generated data from the fit to the MINERvA data with $t_0 =0~\mathrm{GeV}^2$ and the regularization $\lambda=-1$.} \label{figure:pseudodata3}
\end{figure*}

In this Appendix, we study the dependence of the pseudodata analysis on the fit parameter $t_0$. We repeat our study in Section~\ref{sec:deuterium_simulation} for two other choices of the nucleon axial-vector form factor: fits to the pseudodata with $t_0 =-0.28~\mathrm{GeV}^2$ without the regularization and $t_0 =0~\mathrm{GeV}^2$ with the regularization $\lambda=-1$ in Eq.~(\ref{eq:lambda_definition}). We present the corresponding $\chi^2$ and $r_A^2$ in Tables~\ref{tab:pseudodatafits2} and~\ref{tab:pseudodatafits3}, and plot the nucleon axial-vector form factor in Figs.~\ref{figure:pseudodata2}~and~\ref{figure:pseudodata3}. The sensitivity of the squared axial-vector radius $r^2_A$ to the choice of $k_\mathrm{max}$ is reduced as $t_0$ is decreased since the value of $z$ at $Q^{2}=0$ is reduced. At $t_0=0$ in Table~\ref{tab:pseudodatafits3}, $z$ is exactly 0 at $Q^{2}=0$ and therefore the uncertainties for $k_\mathrm{max}=6$ and $7$ are the same. According to Figs.~\ref{figure:pseudodata},~\ref{figure:pseudodata2},~and~\ref{figure:pseudodata3}, the biases in extracting the nucleon axial-vector form factor without a proper implementation of radiative corrections do not change significantly with variation of $t_0$ and various types of the regularization. However, the corresponding small variations largely depend on the (anti)neutrino flux configuration. We also find generally larger $\chi^2$ and larger errors in the nucleon axial-vector radius decreasing the value of $t_0$ in our study, besides the Hyper-K flux configuration that has the lowest energies and can be described with smaller values of $t_0$. For the existent deuterium bubble chamber data sets, we suggest using larger values of the parameter $t_0$.  

\begin{table}[hptb]
    \centering
    \begin{tabular}{l|c|c|c|c}
        &\multicolumn{2}{c|}{$k_\mathrm{max} = 6$} &\multicolumn{2}{c}{$k_\mathrm{max} = 7$} \\  \cline{2-5}  & & & & \\[-2.5ex]
 Flux mode and QED & $\chi^2$ & $\left(r_A^2\right)_\mathrm{true} - r^2_A,~\mathrm{fm}^2$ & $\chi^2$ & $\left(r_A^2\right)_\mathrm{true} - r^2_A,~\mathrm{fm}^2$ \\[0.4ex] \hline \hline
DUNE              &  &  & \\ \hline
\text{RC} fit to \text{RC} data &  57.1 &  0.038(15) &  49.3 &  0.133(29) \\
\text{RC} fit to \text{no RC} data &  54.7 &  -0.089(14) &  51.5 &  -0.034(29) \\
\text{no RC} fit to \text{RC} data &  61.8 &  0.144(17) &  58.5 &  0.198(33) \\
\text{no RC} fit to \text{no RC} data &  44.2 &  0.010(14) &  42.5 &  0.035(28) \\
\hline
Hyper-K           &  &  & \\ \hline
\text{RC} fit to \text{RC} data &  57.3 &  0.064(23) &  71.1 &  -0.039(48) \\
\text{RC} fit to \text{no RC} data &  63.1 &  0.087(24) &  58.7 &  0.173(41) \\
\text{no RC} fit to \text{RC} data &  50.2 &  -0.011(23) &  50.3 &  -0.063(42) \\
\text{no RC} fit to \text{no RC} data &  49.6 &  0.010(22) &  56.0 &  -0.042(43) \\
\hline
MINERvA           &  &  & \\ \hline
\text{RC} fit to \text{RC} data &  59.1 &  0.025(14) &  53.3 &  0.098(28) \\
\text{RC} fit to \text{no RC} data &  66.4 &  -0.168(15) &  65.2 &  -0.166(31) \\
\text{no RC} fit to \text{RC} data &  80.3 &  0.197(17) &  72.3 &  0.287(34) \\
\text{no RC} fit to \text{no RC} data &  48.9 &  -0.001(13) &  46.9 &  0.026(27) \\
\hline
BEBC              &  &  & \\ \hline
\text{RC} fit to \text{RC} data &  96.6 &  0.245(19) &  89.3 &  0.333(38) \\
\text{RC} fit to \text{no RC} data &  47.3 &  0.001(13) &  45.3 &  0.033(27) \\
\text{no RC} fit to \text{RC} data &  93.3 &  0.240(18) &  87.2 &  0.316(38) \\
\text{no RC} fit to \text{no RC} data &  47.4 &  -0.004(13) &  45.6 &  0.015(27) \\
\hline
    \end{tabular}
    \caption{Same as Table~\ref{tab:pseudodatafits} but for the nucleon axial-vector form factor in the generated data from the fit to the MINERvA data with $t_0 =-0.28~\mathrm{GeV}^2$ without regularization.} \label{tab:pseudodatafits2}
\end{table}

\begin{table}[hptb]
    \centering
    \begin{tabular}{l|c|c|c|c}
        &\multicolumn{2}{c|}{$k_\mathrm{max} = 6$} &\multicolumn{2}{c}{$k_\mathrm{max} = 7$} \\  \cline{2-5}  & & & & \\[-2.5ex]
 Flux mode and QED &  $\chi^2$ & $\left(r_A^2\right)_\mathrm{true} - r^2_A,~\mathrm{fm}^2$ &  $\chi^2$ & $\left(r_A^2\right)_\mathrm{true} - r^2_A,~\mathrm{fm}^2$ \\ [0.4ex] \hline \hline
DUNE              &  &  & \\ \hline
\text{RC} fit to \text{RC} data &  116.5 &  -0.121(29) &  71.1 &  0.031(29) \\
\text{RC} fit to \text{no RC} data &  142.1 &  -0.329(32) &  77.2 &  -0.145(30) \\
\text{no RC} fit to \text{RC} data &  81.0 &  0.061(25) &  64.6 &  0.148(28) \\
\text{no RC} fit to \text{no RC} data &  82.2 &  -0.151(25) &  52.1 &  0.031(25) \\
\hline
Hyper-K           &  &  & \\ \hline
\text{RC} fit to \text{RC} data &  71.0 &  0.020(33) &  61.2 &  0.043(30) \\
\text{RC} fit to \text{no RC} data &  83.4 &  0.045(35) &  69.8 &  0.075(32) \\
\text{no RC} fit to \text{RC} data &  56.8 &  -0.066(31) &  50.6 &  -0.051(28) \\
\text{no RC} fit to \text{no RC} data &  62.4 &  -0.043(31) &  52.7 &  -0.022(29) \\
\hline
MINERvA           &  &  & \\ \hline
\text{RC} fit to \text{RC} data &  131.9 &  -0.173(29) &  74.2 &  0.013(29) \\
\text{RC} fit to \text{no RC} data &  169.0 &  -0.479(33) &  88.4 &  -0.260(32) \\
\text{no RC} fit to \text{RC} data &  106.5 &  0.113(27) &  82.1 &  0.232(31) \\
\text{no RC} fit to \text{no RC} data &  101.2 &  -0.196(26) &  59.2 &  -0.041(26) \\
\hline
BEBC              &  &  & \\ \hline
\text{RC} fit to \text{RC} data &  117.4 &  0.179(28) &  96.3 &  0.289(34) \\
\text{RC} fit to \text{no RC} data &  100.0 &  -0.192(26) &  58.0 &  -0.035(26) \\
\text{no RC} fit to \text{RC} data &  113.8 &  0.168(28) &  93.5 &  0.277(33) \\
\text{no RC} fit to \text{no RC} data &  98.1 &  -0.202(25) &  57.2 &  -0.048(26) \\
\hline
    \end{tabular}
    \caption{Same as Table~\ref{tab:pseudodatafits} but for the nucleon axial-vector form factor in the generated data from the fit to the MINERvA data with $t_0 = 0~\mathrm{GeV}^2$ and the regularization $\lambda=-1$.} \label{tab:pseudodatafits3}
\end{table}

\end{document}